\documentclass[aps,prd,floatfix,amsmath,amssymb,preprint,eqsecnum,nofootinbib]{revtex4}
\usepackage{graphicx}
\usepackage{dcolumn}
\usepackage{bm}
\newcommand{\beq}{\begin{equation}}
\newcommand{\eeq}{\end{equation}}
\newcommand{\bea}{\begin{eqnarray}}
\newcommand{\eea}{\end{eqnarray}}
\newcommand{\ba}{\begin{array}{ccc}}
\newcommand{\ea}{\end{array}}
\newcommand{\nn}{\nonumber \\}

\def\bea{\begin{eqnarray}}
\def\eea{\end{eqnarray}}
\DeclareMathOperator{\Tr}{Tr}

\usepackage{setspace} 
\usepackage{amsmath}
\usepackage{amssymb}

\begin{document}
\preprint{PUPT-2419}
\preprint{MAD-TH-12-06}

\title{Strange Metals in One Spatial Dimension}

\author{Rajesh Gopakumar}
\affiliation{Harish-Chandra Research Institute, Chhatnag Road, Jhusi, Allahabad, India 211019}

\author{Akikazu Hashimoto}

\affiliation{Department of Physics, University of Wisconsin, Madison, WI 53706, USA}

\author{Igor R. Klebanov\footnote{On leave from
Department of Physics and Center for Theoretical Science, Princeton University.}}
\affiliation{School of Natural Sciences, Institute for Advanced Study, Princeton, NJ 08540, USA}

\author{Subir Sachdev}
\affiliation{Department of Physics, Harvard University, Cambridge MA
02138, USA}
\affiliation{Instituut-Lorentz for Theoretical Physics, Universiteit Leiden,
P.O. Box 9506, 2300 RA Leiden, The Netherlands}

\author{Kareljan Schoutens}
\affiliation{Institute for Theoretical Physics, University of Amsterdam, Science Park 904, P.O.Box 94485, 1090 GL Amsterdam, The Netherlands}

\date{\today \\
\vspace{.1in}}

\begin{abstract}
We consider $1+1$ dimensional $SU(N)$ gauge theory coupled to a
multiplet of massive Dirac fermions transforming in the adjoint
representation of the gauge group. The only global symmetry of this
theory is a $U(1)$ associated with the conserved Dirac fermion number,
and we study the theory at variable, non-zero densities.  The high
density limit is characterized by a deconfined Fermi surface state
with Fermi wavevector equal to that of free gauge-charged fermions.
Its low energy fluctuations are described by a coset conformal field
theory with central charge $c=(N^2-1)/3$ and an emergent $\mathcal{N}
= (2,2)$ supersymmetry: the $U(1)$ fermion number symmetry becomes
an $R$-symmetry.  We determine the exact scaling dimensions
of the operators associated with Friedel oscillations and pairing
correlations.  For $N>2$, we find that the symmetries allow relevant
perturbations to this state. We discuss aspects of the $N\rightarrow
\infty$ limit, and its possible dual description in $AdS_3$ involving
string theory or higher-spin gauge theory. We also discuss the low
density limit of the theory by computing the low lying bound state
spectrum of the large $N$ gauge theory numerically at zero density,
using discretized light cone quantization.
\end{abstract}

\maketitle

\section{Introduction}
\label{sec:intro}

An important aim of many applications of the AdS/CFT correspondence to
condensed matter physics is the description of quantum matter at
variable, non-zero densities. Here `density' refers to the conserved
charge, $Q$, of a global $U(1)$ symmetry of the underlying quantum
field theory in $d$ spatial dimensions. Our interest here will be
restricted to zero temperature states which do not break translational
symmetry or the global $U(1)$ symmetry.  Thus, we will not consider
`solids,' `charge density waves' or `superfluids.'  In the traditional
phases of condensed matter physics, the only remaining possibilities
for non-zero density states are the Landau Fermi liquid in dimension
$d \geq 2$, and the Luttinger liquid in dimension $d=1$. Both these
states are characterized by a Fermi momentum, $k_F$, whose value obeys
the Luttinger relation: the volume enclosed by the $(d-1)$-dimensional
surface in momentum space at $k_F$ is proportional to the density, $
Q/L^d$, with the same proportionality constant as that for free
fermions ($L$ is the spatial size which we will take to infinity).

Any other realization of quantum matter whose density can be varied
continuously by an applied chemical potential can generically be
referred to as a `strange metal.' A very promising candidate of a
strange metal is a model of fermions at non-zero density coupled to an
Abelian or non-Abelian gauge field. The non-Fermi liquid effects are
strongest in $d=2$, and this model has been the focus of much study in
the condensed matter literature
\cite{pincus,reizer,palee,monien,sonqcd,Schafer1,Schafer2,qcdreview,sslee,metnem,mross,metzner,bartosch}. The
theory scales to strong coupling, and a perturbative expansion in the
gauge coupling constant cannot be used to analyze the leading infrared
behavior. The flavor large $N_f$ expansion also leads to difficulty:
an expansion in the inverse number of fermion flavors cannot be
reduced to counting fermion loops because of infrared divergences
\cite{sslee,metnem}.

Another possible approach is to take the gauge-charged fermions in the
{\em adjoint\/} representation of the gauge group, and to then take
the `t Hooft large $N$ limit for the $SU(N)$ gauge group. In this
case, infrared divergences do not spoil the naive counting in powers
of $N$, and so even the non-zero density case has a $1/N$ expansion
controlled by the genus of the surface defined by a Feynman graph, as
in all matrix models \cite{solvay}.  However, one is then left with
the generally intractable task of summing all graphs with a given
genus. For certain supersymmetric gauge theories, such matrix field
theories can, in principle, be solved in the large $N$ limit by the
AdS/CFT correspondence
\cite{Maldacena:1997re,Gubser:1998bc,Witten:1998qj}.  Studies of such
finite density models by the AdS/CFT correspondence
\cite{nernst,sslee0,hong0,zaanen1,hong1,denef,faulkner,pufuigor,polchinski2,gubserrocha,gubser2,hong2,kiritsis,kiritsis2,kachru2,larus1,larus2,pp}
have so far only provided a rather incomplete picture of the non-zero
density quantum state. The boundary theory has density, $Q/L^d$, which
scales as $N^2$, and essentially all of this density is associated in
the bulk with degrees of freedom which are beyond the infrared
horizon, with an unknown fate. Under appropriate parameter regimes,
gauge-invariant probe fermions (`mesinos') can acquire a Fermi
surface; however such a Fermi surface is only associated with a
density of order unity, and is incidental to the physics of the
non-Fermi liquid state \cite{ssffl,liza,tadashi1,hyper,hyper2,evajoe}.
These probe Fermi surfaces are analogous to conduction electron Fermi
surfaces in the `fractionalized Fermi liquid' state of Kondo lattice
\cite{ffl1,ffl2}, and do not yield much information on the underlying
non-Fermi liquid state.  In certain uncontrolled computations, all of
the boundary density $Q/L^d$ can be associated with visible Fermi
surfaces in the bulk \cite{sean1,sean2,sean3,hong4,ssfl,mcgreevy}, but
then the resulting state is a Fermi liquid, although an interesting
non-Fermi liquid state seems to have been obtained in recent work
\cite{mcgreevy}.

In an attempt to shed light on the difficult question of the fate of
the `hidden' matter of density proportional to $N^2$, this paper will
examine the problem of adjoint Dirac fermions, at non-zero density,
coupled to a $SU(N)$ gauge field in $1+1$ dimensions, i.e. for
$d=1$. We will show that a number of exact results can be obtained for
general $N$, which we hope will help elucidate the structure of the
large $N$ limit of such matrix field theories.

We consider the theory with Lagrangian
\beq
\mathcal{L} = \mbox{Tr} \left[ \bar{\Psi} \left( i \gamma^\mu D_\mu - m - \mu \gamma^0 \right) \Psi \right] - \frac{1}{2 g_{YM}^2} \mbox{Tr}\, F_{\mu\nu} F^{\mu\nu}
\label{L}
\eeq
with a $SU(N)$ gauge field $A_\mu$, gauge field strength $F_{\mu\nu}$,
gauge coupling $g_{YM}$, and adjoint 2-component complex Dirac
fermions $\Psi$ with mass $m$.  The chemical potential $\mu$ couples
to a global $U(1)$ charge which is distinct from all the $SU(N)$ gauge
charges. Note that this $U(1)$ is the {\em only\/} global symmetry of
this Lagrangian. An analogous $d=1$ model was examined earlier for
adjoint Majorana fermions
\cite{igor1,Kutasov:1993gq,igor2,Gross:1995bp,igor3}: in that case
there is no global $U(1)$ that can be coupled to a chemical potential;
it was found that the ground state had an energy gap to all
excitations, even at $m=0$.  As we will see here, just introducing a
global $U(1)$ by making the fermions complex is sufficient to
transform the physics, and a gapless compressible state is obtained
provided $\mu$ is large enough, or when $\mu=m=0$.

This theory is characterized by three energy scales, $m$,
$g_{YM}\sqrt{N}$, and $\mu$. We will consider first the ``high
density'' limit $\mu \gg m,g_{YM}\sqrt{N}$, where we can begin the
analysis with a Fermi sea of free gauge-charged fermions. Next, we
will consider the opposite ``low density'' limit, where $m\ll
g_{YM}\sqrt{N}$ while $\mu$ is comparable with $m$. Here we have to
begin with an analysis of the spectrum of the $SU(N)$ singlet
excitations of the zero density vacuum: this will be carried out via
the discrete light-cone quantization (DLCQ)
\cite{Pauli:1985pv,Hornbostel:1988fb}.

\begin{figure}
  \centering
  \includegraphics[width=2.3in]{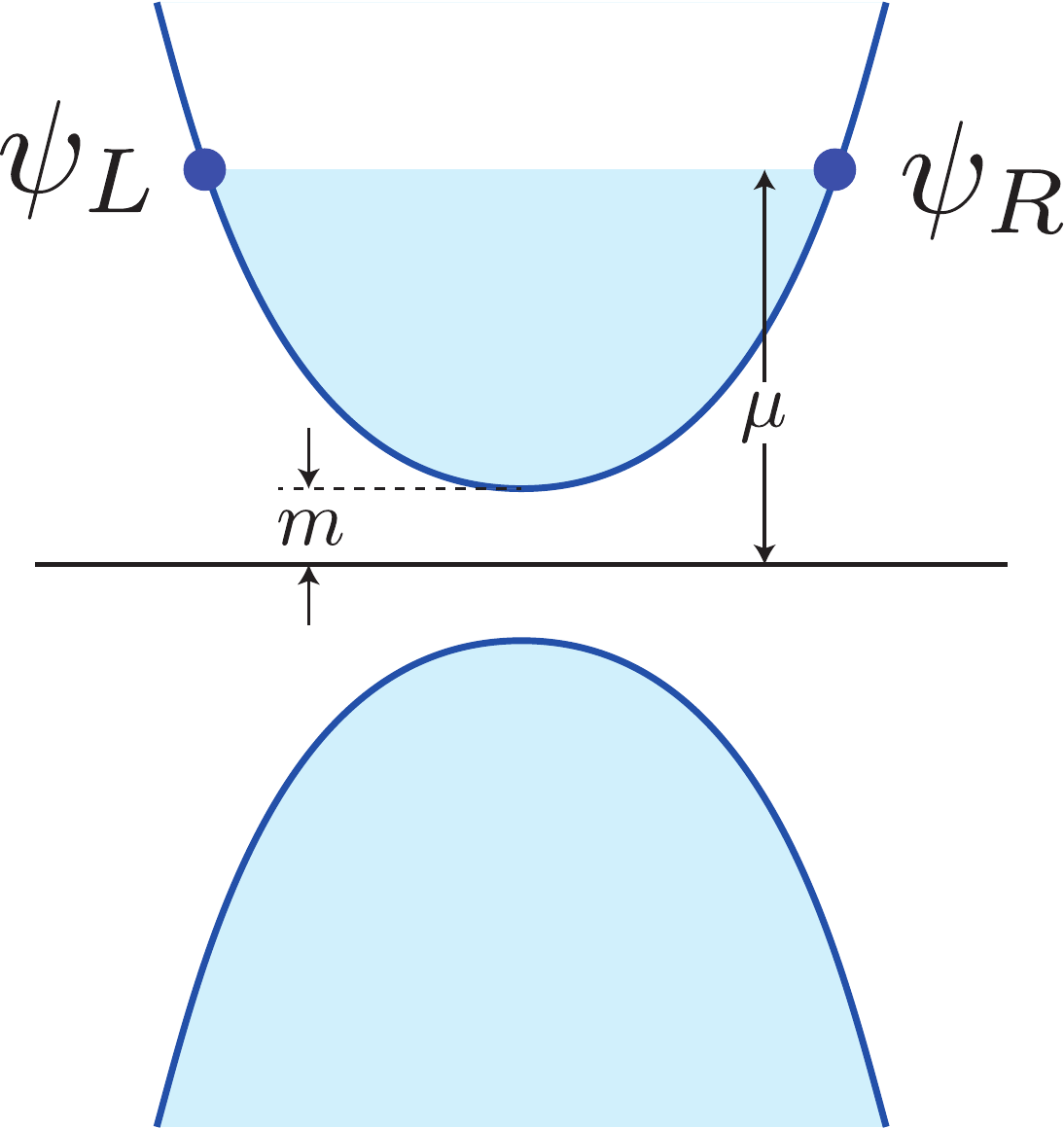}
  \caption{Energy dispersion of the Dirac fermions as a function of momentum. The full line is the zero of energy. The shaded region represents the occupied states. The filled circles are at $\pm k_F$.}
  \label{fig:dispersion}
\end{figure}

We begin with a description of our results for the high density
theory.  The theory of free fermions has a Fermi wavevector related to
the variable $U(1)$ density $Q/L$ by
\beq
\frac{Q}{L} = (N^2 - 1) \frac{k_F}{\pi}\ .  \label{fermi}
\eeq
The $N^2-1$ prefactor is a characteristic signature identifying this
Fermi surface as that of gauge-charged fermions; this Fermi surface is
`hidden' \cite{hyper} because the single fermion Green's function is
not gauge-invariant.  The Luttinger relation implies that this value
of $k_F$ will not be renormalized \cite{oshikawa}. We can analyze the
infrared singular effects of the gauge interactions by writing down a
continuum theory obtained by linearizing the fermions about the Fermi
wavevector. Then, following a procedure standard in the condensed
matter physics literature, we can express the Dirac fermion in terms of its
right-moving and left-moving components at the Fermi surface%
\beq
\Psi (x, t) \sim {1 \over \sqrt{2 E_F}}  \left( u(-k_F) \psi_R (x, t) e^{i k_F x} + u(k_F) \psi_L (x, t) e^{-i k_F x} \right);
\label{psiLR}
\eeq
where $u(\pm k_F)$ are the standard Dirac basis spinors at the Fermi
wavevectors; see (\ref{psiph}) and Appendix~\ref{app:DLCQ} for a more
detailed explanation, and see figure~\ref{fig:dispersion} for an
illustration of this field redefinition.
We will assume that $\psi_{L,R}$, and the gauge-fields, are slowly
varying on the spatial scale $k_F^{-1}$, and so all spatial integrals
of fields multiplied by non-zero integral powers of $e^{\pm i k_F x}$
vanish.  (In condensed matter models, such theories are obtained in
the continuum limit of a lattice Hamiltonian, and in this context we
are assuming that density is incommensurate, and so there is no
`umklapp' scattering.) An immediate consequence is that the resulting
low energy theory has an emergent global $U(1)$ conservation law: the
total number of left-moving and right-moving fermions are {\em
separately\/} conserved. We will denote these two $U(1)$ charges as
$Q_L$ and $Q_R$ respectively, with $Q= Q_L + Q_R$. This $U(1) \times
U(1)$ global symmetry will be crucial to our analysis. All operators
appearing in the effective low-energy Lagrangian must have {\em
both\/} $Q_L = 0$ and $Q_R = 0$.

As we will describe in section~\ref{sec:high}, the high-density, low
energy theory so obtained is a two dimensional conformal field theory
(CFT), associated with the coset
\beq
\frac{SU(N)_N \otimes SU(N)_N}{SU(N)_{2N}} \label{coset}
\eeq
of central charge
\beq \label{centralcharge}
c = \frac{N^2 - 1}{3}.
\eeq
The two dimensions of the CFT are the euclidean continuation of the
original $1+1$ dimensions.  With the requirement that the coset CFT
have a global $U(1) \times U(1)$ symmetry, it actually has the
$\mathcal{N} = (2,2)$ supersymmetry \cite{friedan}. For $N=2,3$ the
central charges are $c=1,8/3$, and then the theories coincide with the
$\mathcal{N} = 2$ superconformal minimal models \cite{vecchia} with
$c=3k/(k+2)$ for $k=1,16$ (in the $k=16$ case we actually find a
certain consistent truncation of the minimal model); the $N\geq 4$
theories were only briefly considered earlier \cite{friedan}. For all
$N$, the $R$-charge symmetry of these $\mathcal{N} = (2,2)$
superconformal field theories (SCFTs) is $U(1) \times U(1)$, and this
provides the needed global symmetry; the SCFT has no other global
flavor symmetries.  Note that this supersymmetry is an emergent
symmetry at low energies and high densities; it is not a symmetry of
the underlying Lagrangian. It is also remarkable that the diagonal
$R$-charge is conjugate to the chemical potential, $\mu$, as has been
assumed by fiat in many earlier higher dimensional studies of non-zero
density quantum matter.

These two dimensional SCFTs are our `strange metals.' They are $T=0$
phases with variable density in models with only a global $U(1)$
symmetry, but with a central charge which can become much greater than
unity. The density fluctuations associated with the $U(1)$ symmetry
cannot be represented by a gapless scalar
field which is decoupled from all other sectors,
as is the case for Luttinger liquids (exceptions for the $N=2$, $c=1$ case
will be discussed in detail below). We note that the `Bose
metal' phases found in multi-leg ladder models in
Refs.~\cite{mishmash} also have $c>1$ and only a global $U(1)$
symmetry, although they are not expected to be described by our SCFTs.

Armed with this construction of the SCFTs, we will compute exact
scaling dimensions of gauge-invariant operators. An important
observable which is sensitive to the presence of the underlying Fermi
surface of the deconfined fermions is the Friedel oscillation in
response to a localized perturbation coupling to the density.  Upon
perturbing the Lagrangian (\ref{L}) via $\mathcal{L}_{\rm imp} =  \mathcal{L} +
\lambda \, \delta(x) \rho(x,t)$, where the density operator
\beq\rho
\equiv \mbox{Tr} (\bar{\Psi} \gamma^0 \Psi ) =\mbox{Tr}\left(\psi_L^\dag \psi_L + \psi_R^\dag  \psi_R + {m\over \mu} e^{- 2 i k_F x} \psi_L^\dag \psi_R  + {m\over \mu} e^{2 i k_F x} \psi_R^\dag \psi_L \right) \ ,
\eeq
the Friedel oscillation response is
\beq
\left\langle \rho (x) \right\rangle_{\rm imp}
\propto \lambda \, \frac{\cos (2 k_F x)}{|x|^{2 \Delta_F}} + \ldots ,\label{friedel}
\eeq
where $\Delta_F$ is the scaling dimension of the operator
$\mbox{Tr}( \psi_L^\dagger \psi_R)$, which we will call the Friedel operator in the CFT.  Equivalently, we can relate the Friedel oscillation to an oscillatory
term in the density-density correlator in the original system without an impurity:
\beq
\left\langle \rho (x) \rho (x') \right\rangle
\propto  \frac{\cos (2 k_F (x-x'))}{|x-x'|^{2 \Delta_F}} + \ldots . \label{friedel2}
\eeq
Our exact result for $\Delta_F$,
obtained by finding the smallest scaling dimension of operators with
$Q_L=1$ and $Q_R=-1$, is
\beq
\Delta_F = {1}/{3}  \quad \mbox{for all $N \geq 2$.} \label{DeltaF}
\eeq
Observation of the oscillatory terms in (\ref{friedel}) and
(\ref{friedel2}) constitutes a measurement of the $k_F$ in
(\ref{fermi}), and is a direct signature of the gauge-charged Fermi
surfaces in these strange metals. Unfortunately, we do not determine
the $N$ dependence of the missing proportionality constants in
(\ref{friedel}) and (\ref{friedel2}); the vanishing of this
proportionality constant in the $N \rightarrow \infty$ limit is the
presumed reason for the absence of such Friedel oscillations in
existing studies via the AdS/CFT correspondence \cite{larus3}.

A second important observable is the fermion pair operator $\mbox{Tr}
( \psi_L \psi_R )$. Condensation of this operator leads to a
superfluid ground state.  The spatial decay of its two-point
correlations is determined by its scaling dimension $\Delta_P$. Our
result for $\Delta_P$ was obtained by finding the smallest scaling
dimension of operators with $Q_L=Q_R=-1$:
\beq
\Delta_P = 1/3   \quad \mbox{for all $N \geq 3$.} \label{DeltaP}
\eeq
For $N=2$ there is no fermion pair operator $\mbox{Tr} ( \psi_L \psi_R
)$ in the CFT, and so in the original gauge theory we expect the two-point functions of
$Q=2$ and $Q=-2$ operators to decay exponentially fast.
Instead, the lowest CFT operator with $Q_L=Q_R$ appears for $N=2$ at $Q_L=Q_R=-3$
and has scaling dimension $3$ (see section \ref{subsec:N=2}).
In Appendix~\ref{app:fund} we review the Luttinger liquid of fermions with short-range interactions ({\em e.g.\/}
the Thirring model at non-zero density), and find that it
obeys $\Delta_F = 1/\Delta_P$;
this identity is clearly not obeyed by the present adjoint matter theory.
Appendix~\ref{app:fund} also computes the values of $\Delta_{F,P}$ in models of fundamental
Dirac fermions coupled to a SU($N$) gauge field.

Finally, to assess the stability of the theory in (\ref{coset}) as a
description of the low energy limit of (\ref{L}), we have to determine
the scaling dimensions of all perturbations allowed by symmetry: these
are all operators with $Q_L = Q_R = 0$.  Here we again find a
distinction between $N=2$ and $N \geq 3$. For $N \geq 3$ the smallest
scaling dimension of such an operator is
\beq
\mbox{dim} \left[ \mbox{Tr} ( \psi_L^\dagger \psi_L \psi_R^\dagger \psi_R ) \right] = \frac{2 (N-2)}{3N},
\label{eq:mostrelevant}
\eeq
which is smaller than 2, and so always relevant. So the $N \geq 3$
SCFT$_2$ is unstable to such a perturbation. We are not able to assess
the $N$ dependence of the coefficient of such a perturbation, or the
ultimate fate of the ground state. A natural conjecture is that this
is an instability to a paired superfluid. In contrast, for $N=2$ there
are no relevant perturbations, and only a marginal perturbation.

The low density limit will be considered in
section~\ref{sec:low}. Here we determine the spectrum of $SU(N)$ singlet
excitations above the zero density vacuum with the aim of using this
as input to describe the finite $\mu$ state as a dilute gas of such
states.  We determine the mass $M$ of the lightest state for a series
of values of $Q$; we will obtain a dilute gas of such states for $\mu
> M/Q$. So we need to determine the value of $Q$ for which $M/Q$ is a
minimum. Our numerical analysis, carried out in the limit $N=\infty$,
suggests that $M/Q$ may accumulate to a dense set of decreasing values
as $Q$ becomes larger (see figure~\ref{extrapolate}). This suggests
that $M/Q$ becomes degenerate at some value in the limit of large $Q$
(however, more extensive numerical work is needed to decide if the
degeneracy is actually present).  This degeneracy would indicate that
even in the low density limit it is not appropriate to use a
description of a dilute gas of gauge-neutral particles, and suggests
the possibility that the gauge-charged-Fermi-sea description of the
high density limit applies even at low densities.

\section{High density}
\label{sec:high}

In this section we will analyze the regime $\mu \gg m, g_{YM}\sqrt{N}$. We
will derive the SCFT$_2$ in (\ref{coset}), and then analyze
its properties in the subsequent subsections.

We begin by writing the Hamiltonian for the free Dirac fermion in
(\ref{L}) in terms of particle, $p$, and hole, $h$, creation and
annihilation operators introduced via (\ref{modexpan}):
\beq
H_0 = \int \frac{dk}{2\pi} \mbox{Tr}  \left[ ( \sqrt{k^2+m^2} - \mu) p^\dagger (k) p(k) -
 ( \sqrt{k^2+m^2} + \mu) h^\dagger (k) h(k)\right] \label{h0}
\eeq
where $k$ is spatial momentum. This defines $k_F$ by $\mu =
\sqrt{k_F^2 + m^2}$. Now we introduce the left and right movers by
\beq
\psi_R (k) = p(k_F + k) \quad, \quad \psi_L (k) = p(-k_F + k), \label{psiph}
\eeq
and then linearize about $k_F$ by approximating
\beq
H_0 = \int \frac{dk}{2\pi} v \, k  \, \mbox{Tr} \left[ \psi_R^\dagger (k) \psi_R (k) - \psi_L^\dagger (k) \psi_L (k) \right]
\eeq
where the velocity $v = k_F /\sqrt{k_F^2 + m^2}$. We will henceforth
set $v=1$.  Carrying out the same mapping to low energy degrees of
freedom in the presence of the gauge field, we obtain the effective
Lagrangian
\bea
\mathcal{L}_{\rm eff} &=& \mbox{Tr} \left[ \psi_R^\dagger ( \partial_\tau -  \partial_x ) \psi_R + \psi_L^\dagger ( \partial_\tau +  \partial_x ) \psi_L
+ (A_\tau - A_x)  [\psi_L^\dagger, \psi_L ] + (A_\tau + A_x)  [\psi_R^\dagger, \psi_R ] \right] \nn
&~&~~- \frac{1}{2 g_{YM}^2} \mbox{Tr}\, F^2 \ . \label{efflag}
\eea
Note that this theory is of the same form as (\ref{L}), but after
setting $\mu=m=0$.  The CFT structure of this theory becomes clearer
upon writing the complex Dirac fields in terms of a pair of Majorana
fields $\psi^a_{L,R}$, and $a=1,2$:
\beq
\psi_{L,R} = \frac{1}{\sqrt{2}} \left( \psi^1_{L,R} + i \psi^2_{L,R} \right) \label{psiphi}
\eeq
and then the Lagrangian becomes
\bea
\mathcal{L}_{\rm eff} &=& \frac{1}{2} \mbox{Tr} \left[ \psi_R^a ( \partial_\tau -  \partial_x ) \psi_R^a + \psi_L^a ( \partial_\tau +  \partial_x ) \psi_L^a + ( A_\tau - A_x)  \psi_L^a \psi_L^a  + ( A_\tau + A_x)  \psi_R^a \psi_R^a  \right] \nn
&~&~~- \frac{1}{2 g_{YM}^2} \mbox{Tr}\, F^2 \ .
\eea
As is well-known, each adjoint Majorana fermion is equivalent to a
$SU(N)$ WZW model at level $N$
\cite{ginsparg,gko,gno,windey1,windey2}, each with central charge
$(N^2 - 1)/2$.  We assume the gauge theory is in the strong coupling
limit ($g_{YM} \rightarrow \infty$), and then the integral over the
gauge field reduces to a constraint: the vanishing of the currents
$J_L = \psi^a_L \psi^a_L$ and $J_R= \psi^a_R \psi^a_R$.  It is easily
verified that these currents obey a $SU(N)$ Kac-Moody algebra at level
$2N$ \cite{kutasov}, and central charge $2(N^2-1)/3$.  The standard
coset construction \cite{gko} then leads to the CFT in (\ref{coset}).

The fact that the low energy  CFT$_2$ of the gauged adjoint
Dirac fermion system has $\mathcal{N} = (2,2)$ supersymmetry was
demonstrated explicitly in (12) of \cite{friedan}. This fact can also
be seen to follow easily from an extension of earlier arguments. It is
useful to write the CFT in (\ref{coset}) in the compact notation
$(N,N;2N)$ as a special case of the general diagonal coset model
$(k,\ell;k+\ell)$ of $SU(N)$, which is $SU(N)_k \otimes SU(N)_\ell
/SU(N)_{k+\ell}$.  Section 3 of \cite{gko} considered the coset model
$(N, \ell; N+\ell)$ and established that it had $\mathcal{N} = (1,1)$
supersymmetry.  We can obtain the second pair of supercharges by
applying the same argument to the coset $(k,N; k+N)$, and so conclude
that the coset $(N,N;2N)$ has $\mathcal{N} = (2,2)$
supersymmetry. This argument also shows that the $R$-charge symmetry
rotates between the two $SU(N)_N$ components i.e.\ between the
two $a$ components of the Majorana fermions.  From (\ref{psiphi}) we
then see that the $R$-charge symmetry is the global $U(1)$ which is
conjugate to the chemical potential.

In the following subsections we will describe the structure of these
theories, including their modular-invariant partition functions and
operator scaling dimensions.

\subsection{$N=2$}
\label{subsec:N=2}

Let us first discuss the simplest non-trivial CFT, corresponding to
the $SU(2)$ gauge theory coupled to an adjoint Dirac fermion. This
$c=1$ CFT may be described by the coset
\beq \label{entwocoset}
\frac{SU(2)_2 \otimes SU(2)_2}{SU(2)_{4}}.
\eeq
The primary fields of this coset theory are therefore labeled by three
$SU(2)$ spins $j_1, j_2, j$ where $j=|j_1-j_2|, \ldots, j_1+
j_2$. Their conformal weights are
\beq \label{cosetdim}
h(j_1, j_2; j)=\frac{j_1(j_1+1)}{4} + \frac{j_2(j_2+1)}{4} -\frac{j(j+1)}{6} + n \ .
\eeq
The ${\cal N}=2$ superconformal symmetry fixes which values of $(j_1,
j_2, j)$ appear in the spectrum. Here $n$ is a non-negative integer
determined in terms of $(j_1, j_2, j)$. It will be zero for the cases
of interest below.

An important quantity characterizing a CFT$_2$ is its modular
invariant partition function on a torus:
\beq \label{modulinv}
Z(\tau,\bar\tau) = \sum_j e^{2\pi i \tau (h_j-c/24)} e^{-2\pi i \bar \tau (\tilde h_j- c/24)}\ ,
\eeq
where the sum runs over the entire spectrum, and $(h_j, \tilde h_j)$
are the holomorphic and anti-holomorphic conformal weights of the
state $j$. Once the modular invariant is known, it is not hard to read
off the spectrum of the theory.  While there is a variety of possible
modular invariants at $c=1$, the ${\cal N}=2$ superconformal
invariance turns out to fix $Z(\tau,\bar \tau)$ completely, up to an
additive constant.

The $c=1$ CFT turns out to be the simplest member of the series of
${\cal N}=2$ superconformal minimal models \cite{vecchia,friedan} with
central charges $c={3k}/({k+2})$: it is its $k=1$ member. The
dimensions of the ${\cal N}=2$ superconformal primary fields, and
their $U(1)$ charges are in general given by \cite{vecchia,friedan}
\beq \label{mindims}
h(p,s,r)= {p^2-1 -(s-r)^2\over 4 (k+2)}+ {|r|\over 8}\ , \qquad q={s-r\over 2 (k+2)}+ {r\over 4}\ ,
\eeq
where $1\leq p\leq k+1$, $|s|\leq p-1$, and $p-s$ must be odd. In the
NS sector $r=0$, while in the R sector $r=\pm 1$.  The NS sector
operators with $p=s+1$ have $h=q$; these are the special operators
that form the chiral ring \cite{lerche}.  We should note that each
primary field of the extended ${\cal N}=2$ algebra gives rise to
various Virasoro primary fields obtained by acting with the
supercharges and $U(1)_R$ current oscillators.

The $k=1$ theory has the following NS-sector ${\cal N}=2$ primaries:
the identity operator, and the operators with $h=\frac 1 6$ and
$U(1)_R$ charge $q=\pm \frac{1}{6}$. In the R sector the primary
fields are $(h=\frac {1} {24}, q=\pm \frac{1} {12})$ and
$(h=\frac{3}{8}, q=\pm \frac {1}{4})$. The former are the R ground
states with $h=c/24$.

Now, let us recall the $c=1$ CFT of a compact massless scalar field
$\phi$ of radius $r$, i.e.\  $\phi$ is identified with $\phi+ 2\pi r$.
The torus partition function of such a theory has the simple explicit
form
\beq \label{modulinv2}
Z_{\rm scalar}(\tau,\bar\tau) = |\eta(\tau)|^{-2} \sum_{n=-\infty}^\infty \sum_{w=-\infty}^\infty e^{\pi i \tau k_L^2} e^{-\pi i \bar \tau k_R^2}
\ .
\eeq
The spectrum of left and right momenta is
\begin{equation}
k_L ={n\over r} - w r/2\ , \qquad  k_R ={n\over r} + w r/2 \ ,
\end{equation}
where $n$ and $w$ are the integer momentum and winding numbers,
respectively.  The left and right dimensions of the Virasoro primary
fields,
\beq
\exp (i k_L \phi_L + i k_R \phi_R)
\ ,
\eeq
are $h=k_L^2/2$, $\tilde h= k_R^2/2$. In addition, for a generic
radius $r$, this CFT has certain primary fields with $k_L=k_R=0$ which
occur for $h=\tilde h= n^2$ where $n$ is an integer. The simplest of
such discrete primary fields is the exactly marginal operator
$\partial \phi\bar \partial \phi$ which changes the radius $r$.

It is well-known that the compact scalar theory at radius $r=2\sqrt 3$
(in units where the self-dual radius is $\sqrt 2$) has ${\cal N}=2$
supersymmetry.  This theory is the bosonization of the above ${\cal
N}=2$ minimal model which provides the correct modular invariant. Let
us consider some of the simplest operators in the bosonized theory and
translate them into the original adjoint fermion language.  The
marginal operator that changes the radius, $\partial \phi \bar\partial \phi$,
corresponds to $J_L(z) J_R (\bar z)$, where
\beq
J_L\propto \Tr (\psi_L^\dagger \psi_L)\ , \qquad J_R\propto \Tr (\psi_R^\dagger \psi_R)
\eeq
are the $U(1)\times U(1)$ currents. We identify the $U(1)\times U(1)$ charges in the $k=1$
${\cal N}=2$ minimal model as $q_L= {k_L\over 2\sqrt 3}$ and $q_R=-{k_R\over 2\sqrt 3}$.
It follows that the relation between $n$ and $w$ in the
compact boson model and the integer charges $Q_L$ and $Q_R$ of the
fermions in the gauge theory is
\beq
\label{chargedict}
n= Q_L- Q_R\ , \qquad w= -(Q_L+Q_R)/6
\ ,
\eeq
where $Q_L=6 q_L$ and $Q_R=6 q_R$.

For $n=\pm 1, w=0$ we get $h=\tilde h={1\over 24}$. This corresponds
to $h(\frac 1 2, \frac 1 2; 1)$ in (\ref{cosetdim}).  These two spin
zero operators are products of the R-sector ${\cal N}=2$
superconformal primaries with $h=\frac {1} {24}$.

For $n=\pm 2, w=0$ we get $h_{n=\pm 2} =\tilde h_{n=\pm 2}={1\over 6}$
corresponding to $h(1,0;1)$ or $h(0,1;1)$. These two spin zero
operators are products of the NS-sector ${\cal N}=2$ superconformal
primaries with dimension $\frac {1} {6}$.  In the gauge theory these
operators are $\Tr (\psi_L^\dagger \psi_R)$ with charges $Q_L=1,
Q_R=-1$, and $\Tr (\psi_L \psi_R^\dagger)$ with charges $Q_L=-1,
Q_R=1$.  Their sum is simply the fermion mass term.  The total
dimension of these operators is $\Delta_F=h+\tilde h={1\over 3}$; this
is the scaling exponent for decay of the Friedel oscillations.

For $(n,w)=\pm (3,0)$ we find $h=\tilde h= \frac{3}{8}$. The
holomorphic part of this operator is the $(h=\frac{3}{8}, q=\pm \frac
{1}{4})$ ${\cal N}=2$ superconformal primary from the R sector (in the
coset theory, it is $h(\frac 1 2, \frac 1 2; 0)=\frac 3 8$).

We could also consider $n=\pm 4, w=0$ operators with $h_{n=\pm
4}=\tilde h_{\pm 4}={2\over 3}$. These operators are not
superconformal primary fields, but they are Virasoro primary.  We note
that $h_{\pm 4}= h_{\pm 2}+{1\over 2}$. This suggests that the
$n=\pm 4$ operators are obtained from $n=\pm 2$ by acting with a
holomorphic and an anti-holomorphic supercharge.

For $n=0, w=\pm 1$ we get $h=\tilde h= {3\over 2}$.  According to
(\ref{chargedict}), these are the operators with $Q_L=Q_R=\pm 3$
responsible for superconducting correlations.  Since $e^{i\sqrt
3\phi_L}$ is the supercurrent, we identify the $n=0, w=\pm 1$
operators with double-trace operators which are products of two
super-currents, each having dimension $3/2$.  Their net dimension is
$3$. We do not find, therefore, separate fermion pair
operators $\Tr (\psi_L \psi_R)$ and $\Tr (\psi_L^\dagger
\psi_R^\dagger)$. This is a special feature of the $N=2$ case; we will
see that such distinct operators appear for $N\geq 3$.

\subsection{$N=3$}

In appendix \ref{app:modinv} we outline a general approach to obtaining
the modular invariant partition sum for the $(N,N;2N)$ cosets and,
thereby, the operator content of the theory. We first show how for
$N=2$ this approach reproduces the result given in (\ref{modulinv2})
and then proceed to the case $N=3$.

The proper starting point for the $N=2$ gauged fermion model is the
$SO(6)_1$ invariant partition sum, broken down to $SO(3)_1 \times
SO(3)_1$,
\bea
&&
Z^{SO(6)_1} = |\chi_{\rm 1}^{\rm SO(6)_1}|^2 + |\chi_{\rm v}^{\rm SO(6)_1}|^2+2|\chi_{\rm sp}^{\rm SO(6)_1}|^2
\nn &&
=
|\chi^{SO(3)_1} _{\rm 1}\chi^{SO(3)_1} _{\rm 1} + \chi^{SO(3)_1} _{\rm v} \chi^{SO(3)_1} _{\rm v}|^2 + 4|\chi^{SO(3)_1} _{\rm 1}\chi^{SO(3)_1} _{\rm v}|^2
+ 2 |\chi^{SO(3)_1} _{\rm sp} \chi^{SO(3)_1} _{\rm sp}|^2 \ .
\eea

We consider the branching rules of the relevant combinations of
$SO(3)_1$ characters.  The ${\cal N}=(2,2)$ superconformal symmetry of
the coset CFT$_2$ guarantees that chiral branching functions will
organize into characters of ${\cal N}=2$ SCFT, defined as
\beq
{\rm ch}^{\rm R,NS}_{h}= \Tr[e^{2\pi i \tau (L_0-c/24)}]
\eeq
(the complete characters of the ${\cal N}=2$ superconformal symmetry
would keep track the $U(1)$ charges as well - for simplicity we
suppress those in our notations).  Considering the explicit leading
terms in the various characters, we established the following
relations
\bea
\lefteqn{\chi^{SO(3)_1}_{\rm 1} \chi^{SO(3)_1}_{\rm 1} + \chi^{SO(3)_1}_{\rm v} \chi^{SO(3)_1}_{\rm v} + 2 \chi^{SO(3)_1}_{\rm 1} \chi^{SO(3)_1}_{\rm v}}
\nn
&& \quad
= {\rm ch}^{\rm NS}_0 \, (\chi^{SU(2)_4}_{(0)} +  \chi^{SU(2)_4}_{(4)})
  + 2 \, {\rm ch}^{\rm NS}_{1/6} \, \chi^{SU(2)_4}_{(2)}
\nonumber \\[2mm]
\lefteqn{\chi^{SO(3)_1}_{\rm 1} \chi^{SO(3)_1}_{\rm 1} + \chi^{SO(3)_1}_{\rm v} \chi^{SO(3)_1}_{\rm v} - 2 \chi^{SO(3)_1}_{\rm 1} \chi^{SO(3)_1}_{\rm v}}
\nn
&& \quad
= \widetilde{\rm ch}^{\rm NS}_0 \, (\chi^{SU(2)_4}_{(0)} +  \chi^{SU(2)_4}_{(4)})
  + 2 \, \widetilde{\rm ch}^{\rm NS}_{1/6} \, \chi^{SU(2)_4}_{(2)}
\nonumber \\[2mm]
\lefteqn{\chi^{SO(3)_1}_{\rm sp} \chi^{SO(3)_1}_{\rm sp} }
 \nn
 && \quad = {\rm ch}^{\rm R}_{1/24} \, (\chi^{SU(2)_4}_{(0)} +  \chi^{SU(2)_4}_{(4)})
 + {\rm ch}^{\rm R}_{3/8} \, \chi^{SU(2)_4}_{(2)} \ .
\eea
Note that the characters $\widetilde{\rm ch}^{\rm NS}$ are obtained by
inserting a $(-1)^F$, with $F$ the fermion parity operator. In the
R-sector the characters $\widetilde{\rm ch}^{\rm R}$ are constants
known as the Witten index of the sector.

For $SU(2)_4$ there exists an exceptional invariant (labeled as $D_4$
in the literature \cite{cappelli}), which groups the characters
according to the automorphism (\ref{eq:auto})
\beq
Z^{SU(2)_4} = |\chi^{SU(2)_4}_{(0)}+ \chi^{SU(2)_4}_{(4)}|^2+ 2 |\chi^{SU(4)_4}_{(2)}|^2 \ .
\eeq
This invariant sets a modular invariant metric on the $SU(2)_4$
characters. Using the metric we can project out the level-4 characters
in the various product terms in the $SO(6)_1$ partition sum. The final
result is the following modular invariant coset partition sum
\bea
Z^{\rm coset (2,2;4)} &=&
{1 \over 2} [ | {\rm ch}^{\rm NS}_{0} |^2 +  2 | {\rm ch}^{\rm NS}_{1/6} |^2
+({\rm ch}^{\rm NS} \to \widetilde{\rm ch}^{\rm NS}) ]
\nn &&
+ {1 \over 2} [ | {\rm ch}^{\rm R}_{3/8} |^2 +  2 | {\rm ch}^{\rm R}_{1/24} |^2] \ .
\eea
Comparing with the expression (\ref{modulinv2}), evaluated at the
${\cal N}=2$ radius $r=2\sqrt 3$, one checks that the two expressions
agree up to a constant which we can write as the sum of the R-sector
Witten indices and which is by itself modular invariant
\beq
Z^{\rm coset (2,2;4)}
= Z_{\rm scalar}- {1 \over 2} \left[ 2 | \widetilde {\rm ch}^{\rm R}_{1/24} |^2 \right]
=Z_{\rm scalar} - 1  \ .
\eeq

We are now ready to take on the case $N=3$. The central charge $c=8/3$
corresponds to the $k=16$ entry in the minimal series of unitary
${\cal N}=(2,2)$ SCFT$_2$. One thus expects that the partition sum can
be expressed as a finite sum of terms of the form ${\rm ch}^{{\cal
N}=2}_{h,q} \overline{\rm ch}^{{\cal N}=2}_{h',q'}$.  Modular
invariant partition sums of this type have been completely classified
\cite{gray} - our challenge is thus to identify the correct entry from
the (rather extensive) list.

The $N=3$ strange metal starts from 16 fermions with partition sum
\bea
&&
Z^{SO(16)_1} = |\chi_{\rm 1}^{\rm SO(16)_1}|^2 + |\chi_{\rm v}^{\rm SO(16)_1}|^2+2|\chi_{\rm sp}^{\rm SO(16)_1}|^2
\nn &&
=
|\chi^{SO(8)_1} _{\rm 1}\chi^{SO(8)_1} _{\rm 1} + \chi^{SO(8)_1} _{\rm v} \chi^{SO(8)_1} _{\rm v}|^2 + 4|\chi^{SO(8)_1} _{\rm 1}\chi^{SO(8)_1} _{\rm v}|^2
+ 8 |\chi^{SO(8)_1} _{\rm sp} \chi^{SO(8)_1} _{\rm sp}|^2 \ .
\eea
A curiosity specific to $N=3$ is that the vector representation of
$SO(8)$ is isomorphic to the spinors - in the coset theory this leads
to degeneracies between NS and R sector dimensions.

As before we now study the branching into characters of $SU(3)_6$
times branching functions which we expect to be characters of a
$W$-algebra extension of ${\cal N}=(2,2)$ superconformal
symmetry.\footnote{See \cite{bouwknegt3} for a general review of
$W$-symmetry and \cite{ahn,schoutens} for $W$-extensions of ${\cal
N}=1$ superconformal symmetry in general cosets involving a $SU(N)_N$
factor.} Working through explicit details, one observes that in the
r.h.s. of the branching rules, the $SU(3)_6$ characters consistently
show up in the combinations
\bea
& \chi^{SU(3)_6}_{(00)} + \chi^{SU(3)_6}_{(60)} + \chi^{SU(3)_6}_{(06)},
& \chi^{SU(3)_6}_{(11)} + \chi^{SU(3)_6}_{(41)} + \chi^{SU(3)_6}_{(14)},
\nonumber \\[2mm]
& \chi^{SU(3)_6}_{(33)} + \chi^{SU(3)_6}_{(30)} + \chi^{SU(3)_6}_{(03)}, & \chi^{SU(3)_6}_{(22)} \ .
\eea
A modular invariant projector for these terms is provided by the ${\cal D}_6$ invariant of $SU(3)_6$ \cite{gannon}
\beq
Z^{SU(3)_6}
 = |\chi_{(00)}+\chi_{(60)}+\chi_{(06)}|^2 +  |\chi_{(11)}+\chi_{(41)}+\chi_{(14)}|^2 +  |\chi_{(33)}+\chi_{(30)}+\chi_{(03)}|^2
   + 3  |\chi_{(22)}|^2 \ .
\eeq

Completing this analysis, we have identified (up to a constant) the
partition sum for the $(3,3;6)$ strange metal with the exceptional
invariant of ${\cal N}=(2,2)$ superconformal symmetry at $c=8/3$ which
in the classification \cite{gray} is labeled as $\widetilde{M}^{4,2}$
with parameters $v=3$, $z=1$, $x=1$. This invariant involves a subset
of all NS and R characters at $k=16$; furthermore, those that survive
are grouped into groups of 6 (4 groups in each sector) and per sector
one group of 3. So in total, 54 fields survive. In the NS sector the
extended characters are (${\cal N}=2$ characters labeled as ${\rm
ch}_{l=p-1,s}$ with $r=0$ for NS and $r=-1$ for R)
\bea
\label{su3ns}
&&
{\rm ch}^{\rm NS,ext}_0 ={\rm ch}^{\rm NS}_{0,0}+{\rm ch}^{\rm NS}_{16,0}+{\rm ch}^{\rm NS}_{16,6}
+{\rm ch}^{\rm NS}_{16,-6}+{\rm ch}^{\rm NS}_{16,12}+{\rm ch}^{NS}_{16,-12}
\nn
&&
{\rm ch}^{\rm NS,ext}_{1/9}={\rm ch}^{\rm NS}_{2,0}+{\rm ch}^{\rm NS}_{14,0}+{\rm ch}^{\rm NS}_{14,6}
+{\rm ch}^{\rm NS}_{14,-6}+{\rm ch}^{\rm NS}_{14,12}+{\rm ch}^{\rm NS}_{14,-12}
\nn
&&
{\rm ch}^{\rm NS,ext}_{1/3} ={\rm ch}^{\rm NS}_{4,0}+{\rm ch}^{\rm NS}_{12,0}+{\rm ch}^{\rm NS}_{12,6}
+{\rm ch}^{\rm NS}_{12,-6}+{\rm ch}^{\rm NS}_{12,12}+{\rm ch}^{\rm NS}_{12,-12}
\nn
&&
{\rm ch}^{\rm NS,ext}_{1/6} ={\rm ch}^{\rm NS}_{6,0}+{\rm ch}^{\rm NS}_{10,0}+{\rm ch}^{\rm NS}_{6,6}
+{\rm ch}^{\rm NS}_{6,-6}+{\rm ch}^{\rm NS}_{10,6}+{\rm ch}^{\rm NS}_{10,-6}
\nn
&&
{\rm ch}^{\rm NS,ext}_{11/18} ={\rm ch}^{\rm NS}_{8,0}+{\rm ch}^{\rm NS}_{8,6}+{\rm ch}^{\rm NS}_{8,-6}
\eea
and in the R sector
\bea
&&
{\rm ch}^{\rm R,ext}_1 ={\rm ch}^{\rm R}_{16,-16}+{\rm ch}^{\rm R}_{16,14}+{\rm ch}^{\rm R}_{16,-10}
+{\rm ch}^{\rm R}_{16,8}+{\rm ch}^{\rm R}_{16,-4}+{\rm ch}^{\rm R}_{16,2}
\nn
&&
{\rm ch}^{\rm R,ext}_{1/9}={\rm ch}^{\rm R}_{14,14}+{\rm ch}^{\rm R}_{14,-10}+{\rm ch}^{\rm R}_{14,8}
+{\rm ch}^{\rm R}_{14,-4}+{\rm ch}^{\rm R}_{14,2}+{\rm ch}^{\rm R}_{2,2}
\nn
&&
{\rm ch}^{\rm R,ext}_{1/3} ={\rm ch}^{\rm R}_{12,-10}+{\rm ch}^{\rm R}_{12,8}+{\rm ch}^{\rm R}_{12,-4}
+{\rm ch}^{\rm R}_{12,2}+{\rm ch}^{\rm R}_{4,-4}+{\rm ch}^{\rm R}_{4,2}
\nn
&&
{\rm ch}^{\rm R,ext}_{2/3} ={\rm ch}^{\rm R}_{10,-10}+{\rm ch}^{\rm R}_{10,8}+{\rm ch}^{\rm R}_{10,-4}
+{\rm ch}^{\rm R}_{10,2}+{\rm ch}^{\rm R}_{6,-4}+{\rm ch}^{\rm R}_{6,2}
\nn
&&
{\rm ch}^{\rm R,ext}_{1/9'} ={\rm ch}^{\rm R}_{8,-4}+{\rm ch}^{\rm R}_{8,2}+{\rm ch}^{\rm R}_{8,8} \ .
\eea
The extended vacuum character has the following content (returning to
the notation ${\rm ch}_{h,q}$ for the ${\cal N}=2$ characters)
\bea
\label{su3vac}
\lefteqn{{\rm ch}^{\rm NS,ext}_0=}
\\
&&
{\rm ch}^{\rm NS}_{h=0,q=0} +{\rm ch}^{\rm NS}_{h=2,q=1/3}+{\rm ch}^{NS}_{h=2,q=-1/3}
+{\rm ch}^{\rm NS}_{h=7/2,q=1/6} +{\rm ch}^{\rm NS}_{h=7/2,q=-1/6} +{\rm ch}^{\rm NS}_{h=4,q=0} \ .
\nonumber
\eea
These fields span a $W$-algebra extension of  ${\cal N}=2$ superconformal symmetry, with extra currents of
dimension $2$, $2$, $7/2$, $7/2$, $4$. The $W$-currents at $h=2$ are given by
\beq
\label{chgspintwo}
W_L={\rm Tr}(\psi_L^\dagger \psi_L^3)\ ,
\qquad W^\dagger_L={\rm Tr}(\psi_L \psi_L^{\dagger 3}) \ .
\eeq
These currents exist for all $N\geq 3$. For $N=2$ the triple products $\psi_L^3$ and $\psi^{\dagger 3}_L$
are $SU(2)$ singlets and the trace vanishes. For $N\geq 3$ explicit expressions for $W_L$ and
$W^\dagger_L$ in terms of component fields $(\psi_L)_A$ and $(\psi_L^\dagger)_A$, with $A=1,2,\ldots N^2-1$
an adjoint index, involve the 3-index $d$-symbols $d_{ABC}$.

The $\widetilde{M}^{4,2}$ invariant reads
\bea
\lefteqn{Z^{\widetilde{M}^{4,2}} =
{1 \over 2} \left[ \sum_{h=0,1/9,1/6,1/3}  | {\rm ch}^{\rm NS,ext}_{h}|^2 + 2  | {\rm ch}^{\rm NS,ext}_{11/18}|^2
  +({\rm ch}^{\rm NS} \to \widetilde{\rm ch}^{\rm NS}) \right. }
 \nn &&  \qquad \qquad
  \left. + \sum_{h=1/9,1/3, 2/3,1}  | {\rm ch}^{\rm R,ext}_{h}|^2 + 2  | {\rm ch}^{\rm R,ext}_{1/9'}|^2 \right] \ .
 \eea
 and the claim is
 \beq
 Z^{\rm coset (3,3;6)} = Z^{\widetilde{M}^{4,2}} - 1/3 \ .
 \eeq
The $U(1)_R$ charges of the fields surviving in the partition sum are
easily extracted from the field labels in the partition sum. In both
the NS and the R sectors, the $U(1)_R$ charges are multiples of $\pm
1/6$.

Comparing with $N=2$, we observe, at $N=3$, the presence of fermion pair
operators ${\rm Tr}(\psi_L\psi_R)$ and ${\rm
Tr}(\psi^\dagger_L\psi^\dagger_R)$ which were absent from the operator spectrum
for $N=2$.
From (\ref{su3ns}) we read off that these operators are in the same
extended symmetry multiplet as the mass (Friedel) operators ${\rm Tr}(\psi_L^{\dagger}\psi_R)$ and ${\rm Tr}(\psi_L\psi_R^{\dagger})$. Both these sets of operators have scaling dimension
$1/3$ - the degeneracy is due to the fact that the zero mode of the charged spin two currents in (\ref{chgspintwo}) relates the two sets of operators.

We also observe the presence of a number of charge
neutral relevant operators.  The most relevant operator is the ($l=2,
s=0$) field in the NS sector, with scaling dimension $\Delta=2/9$, as
in (\ref{eq:mostrelevant}).

 \begin{center}
 \begin{table}
 \begin{tabular}{lccl}
 \hline
 \hline
 current & $(q_L,q_R)$ \qquad & $(h,\tilde h)$ \quad & \\[1mm]
 \hline
 $ J_L \propto {\rm Tr}(\psi_L \psi^\dagger_L)$  & $(0,0)$ & $(1,0)$ & $N\geq 2$
 \\[2mm]
 $G_L \propto {\rm Tr}(\psi_L^3)$ & $(-{1\over 2},0)$ & $({3 \over 2},0)$ & $N\geq 2$
 \\[2mm]
 $G^\dagger_L \propto {\rm Tr}(\psi^{\dagger 3}_L)$ & $({1 \over 2},0)$ & $({3 \over 2},0)$ & $N\geq 2$
 \\[2mm]
 $T_L \propto {\rm Tr}(\psi_L \partial \psi^\dagger_L+\psi_L^\dagger \partial \psi_L)$ & $(0,0)$ & $(2,0)$ & $N\geq 2$
 \\[2mm]
 $W_L \propto {\rm Tr}(\psi_L^\dagger \psi_L^3)$ & $(-{1\over 3},0)$ & $(2,0)$ & $N\geq 3$
 \\[2mm]
$W^\dagger_L \propto {\rm Tr}(\psi_L \psi_L^{\dagger 3})$ & $({1\over3},0)$ & $(2,0)$ & $N\geq 3$
\\[2mm]
 $\ldots$
\\[1mm]
 \hline
 \hline
 \end{tabular}
 \caption{This table lists some of the (left) chiral fields (currents) in the $(N,N;2N)$ coset model. The
currents ${J_L,G_L,G^\dagger_L,T_L}$ constitute an ${\cal N}=2$ superconformal algebra. The
dimension-2 currents $W_L$, $W^\dagger_L$ are the first of an extensive set of extra primary
symmetry generators that exist for $N\geq 3$. }
\label{tab:currents}
 \end{table}
 \end{center}

 \begin{center}
\begin{table}
 \begin{tabular}{lcccc}
 \hline
 \hline
 operator type & $(q_L,q_R)$ & $(h,\tilde h)$ & channel &  \\[1mm]
 \hline
$ {\rm Tr}(\psi_L^\dagger \psi_R)$  & $({1 \over 6}, -{1 \over 6})$ & $({1 \over 6},{1 \over 6})$ & $(10\ldots 01)$ & $N\geq 2$
\\[2mm]
$ {\rm Tr}(\psi_L \psi_R^\dagger)$  & $(-{1\over 6} , {1 \over 6})$ & $({1 \over 6},{1 \over 6})$ & $(10\ldots 01)$ & $N\geq 2$
\\[2mm]
$ {\rm Tr}(\psi_L  \psi_R)$  & $(-{1 \over 6}, -{1\over 6})$ & $({1 \over6},{1 \over 6})$ & $(10\ldots 01)$ & $N\geq 3$
\\[2mm]
$ {\rm Tr}(\psi_L^\dagger  \psi_R^\dagger)$  & $({1 \over 6}, {1 \over 6})$ & $({1 \over 6},{1\over 6})$ & $(10\ldots 01)$ & $N\geq 3$
\\[1mm]
\hline
 ${\rm Tr} (\psi_L \psi_L \psi_R \psi_R)$ & $(-{1 \over 3},-{1 \over 3})$ &
 $({1 \over 3},{1\over 3})$ & $(20\ldots 10)$ & $N\geq 3$
\\[2mm]
${\rm Tr} (\psi_L \psi_L \psi^\dagger_R \psi_R)$ & $(-{1\over 3},0)$ &
 $({1 \over 3},{1\over 3})$ & $(20\ldots 10)$ & $N\geq 3$
\\[2mm]
${\rm Tr} (\psi_L \psi_L \psi^\dagger_R \psi^\dagger_R)$ & $(-{1 \over 3},{1 \over 3})$ &
 $({1 \over 3},{1\over 3})$ & $(20\ldots 10)$ & $N\geq 3$
\\[2mm]
${\rm Tr} (\psi^\dagger_L \psi_L \psi_R \psi_R)$ & $(0,-{1 \over 3})$ &
 $({1 \over 3},{1\over 3})$ & $(20\ldots 10)$ & $N\geq 3$
\\[2mm]
${\rm Tr} (\psi^\dagger_L \psi_L \psi^\dagger_R \psi_R)$ & $(0,0)$ &
 $({1 \over 3},{1\over 3})$ & $(20\ldots 10)$ & $N\geq 3$
\\[2mm]
${\rm Tr} (\psi^\dagger_L \psi_L \psi^\dagger_R \psi^\dagger_R)$ & $(0,{1 \over 3})$ &
 $({1 \over 3},{1\over 3})$ & $(20\ldots 10)$ & $N\geq 3$
\\[2mm]
${\rm Tr} (\psi^\dagger_L \psi^\dagger_L \psi_R \psi_R)$ & $({1 \over 3},-{1 \over 3})$ &
 $({1 \over 3},{1\over 3})$ & $(20\ldots 10)$ & $N\geq 3$
\\[2mm]
${\rm Tr} (\psi^\dagger_L \psi^\dagger_L \psi^\dagger_R \psi_R)$ & $({1 \over 3},0)$ &
 $({1 \over 3},{1\over 3})$ & $(20\ldots 10)$ & $N\geq 3$
\\[2mm]
${\rm Tr} (\psi^\dagger_L \psi^\dagger_L \psi^\dagger_R \psi^\dagger_R)$ & $({1 \over 3},{1 \over 3})$ &
 $({1 \over 3},{1\over 3})$ & $(20\ldots 10)$ & $N\geq 3$
\\[2mm]
same 9 terms & &
 $({1 \over 3},{1\over 3})$ & $(010\ldots 02)$ & $N\geq 3$
 \\[2mm]
same 9 terms & &
 $({2 \over 3},{2\over 3})$ & $(10\ldots 01)$ & $N\geq 3$
\\[1mm]
\hline
 ${\rm Tr} (\psi_L^\dagger \psi_L \psi_R^\dagger \psi_R)$ & $(0,0)$ &
 $({N-2\over 3N},{N-2\over 3N})$ & $(20\ldots 02)$ & $N\geq 3$
\\[2mm]
 ${\rm Tr} (\psi_L^\dagger \psi_L \psi_R^\dagger \psi_R)$ & $(0,0)$ &
 $({N+2\over 3N},{N+2\over 3N})$ & $(010 \ldots 010)$ & $N\geq 4$
  \\[1mm]
 \hline
$ {\rm Tr}(\psi_L^\dagger \psi_L^2 \psi_R \psi_R^{\dagger\, 2})$  & $(-{1 \over 6},{1 \over 6})$ & $({N-2 \over 2N},{N-2 \over 2N})$
& $(30\ldots 011)$ & $N\geq 4$
 \\[2mm]
 $\ldots$
 \\[1mm]
 \hline
$ {\rm Tr}(\psi_L^\dagger \psi_L  \psi_R^\dagger \psi_R)^2$  & $(0,0)$ & $({2(N-2) \over 3N},{2(N-2) \over 3N})$
& $(210\ldots 012)$ & $N\geq 4$
\\[2mm]
$ {\rm Tr}(\psi_L^\dagger \psi_L \psi_R^\dagger \psi_R )^2$  & $(0,0)$ & $({2(N-3) \over 3N},{2(N-3) \over 3N})$
& $(40\ldots 020)$ & $N\geq 4$
 \\[2mm]
 $\ldots$
 \\[1mm]
 \hline
 ${\rm Tr} (\psi_L^\dagger \psi_L \psi_R^\dagger \psi_R)^n$ & $(0,0)$ &
 $(n {N-1-n\over 3N},n {N-1-n\over 3N})$ & $(2n 0\ldots 0 2n)$ &  $N\geq n+2$
 \\[2mm]
 $\ldots$
 \\[1mm]
 \hline
 \hline
 \end{tabular}
 \caption{Operators in the $(N,N;2N)$ coset model. The operators
 listed (all in the NS sector) are primaries with respect to the
 ${\cal N}=(2,2)$ superconformal algebra. The channel indicated is the
 $SU(N)$ representation of the gauged diagonal $SU(N)_{2N}$
 subalgebra. We give the complete list of 2- and 4-fermion primaries
 and in addition list some of the higher operators.}
\label{tab:primaries}
\end{table}
\end{center}

\subsection{General $N$}

We have also obtained several explicit results for $N=4$ and higher,
confirming the general structure outlined above and in appendix
\ref{app:modinv}.  We defer the details of the description of the
general $N$ case to a forthcoming publication. Here we briefly
indicate some of the findings, paying special attention to the
extrapolation to large $N$.

\begin{itemize}

\item
In tables \ref{tab:currents} and \ref{tab:primaries} we list the
general form of the leading currents and primary field operators in
the $(N,N;2N)$ coset. These operators are all in the NS sector.  We
see that their scaling dimensions are either $N$-independent or are
such that $h$ ($\tilde{h}$) converge in the large $N$ limit to
$n_{L(R)}/6$ if the operator involves $n_{L(R)}$ left (right) moving
fermions. In particular, we see that both fermion pair operators and
the Friedel operators are present in the spectrum and have a scaling
dimension $\Delta_F=\Delta_P=1/3$ independent of $N$. Their degeneracy
is due to the presence of the charged spin two currents
(\ref{chgspintwo}) which are part of the extended $W$-symmetry for any
$N\geq 3$.

\item
One can also quickly verify that the scaling
dimensions in the R sector satisfy
\beq
h^{\rm R} \geq {c \over 24} = {N^2-1 \over 72}
\eeq
so that for large $N$ all R-sector operators are irrelevant.

\item
The algebraic structure of the general-$N$ coset appears to be highly involved. The cosets of the form
\beq
\label{wnmin}
\frac{SU(N)_k \otimes SU(N)_1}{SU(N)_{k+1}}
\eeq
are known to have an extended $W_N$ symmetry with one chiral current for each spin $s=2,\ldots N$ \cite{bais, bouwknegt3}.
For a general diagonal coset (which includes our coset as a special case)
\beq
\label{wngen}
\frac{SU(N)_k \otimes SU(N)_{\ell}}{SU(N)_{k+{\ell}}}
\eeq
we can construct higher spin-$s$ currents from polynomial (of degree
$s$) combinations of the individual numerator $SU(N)$ currents which
commute with the diagonal $SU(N)$. See (7.42) and (7.43) of
\cite{bouwknegt3} for an explicit expression for the spin-3 current.
These general cosets are expected to have many additional currents as
well.  In the case of $N=3$ we found two charged
spin two fields, see (\ref{chgspintwo}), in addition to the
stress tensor. Since these are primaries of the ${\cal N}=2$
supersymmetry, this then implies the existence of charged spin
$\frac{5}{2}$ and spin-3 currents as well. Similarly, we see from
(\ref{su3vac}) that there are charged spin $\frac{7}{2}$ currents and
their ${\cal N}=2$ descendants.

For general $N$ we expect to be able to build many chiral operators
from $\psi_L$, $\psi_L^{\dagger}$ and holomorphic derivatives.  In the
large $N$ limit we do not expect any trace relations for such
operators of spin $s \ll N$. Then by the usual counting arguments for
words built from matrix valued fields one expects to roughly see a
Hagedorn growth in the number of these currents (as a function of the
dimension, which is the same as the spin). Thus we might expect an
algebra much larger than the conventional $W_N$ symmetry algebra. It
would be interesting to work out the consequences of this larger
symmetry algebra.

\end{itemize}

There are a number of issues that would be important to explore
further for this class of coset models.

A key algebraic structure of ${\cal N}=2$ SCFT$_2$ is the chiral ring:
the collection of NS sector primaries with $q=\pm h$, which form a
closed algebra under fusion \cite{lerche}. In the $N=3$ case, we read
off from the NS sector primaries in (\ref{su3ns}) that the only chiral
primaries are those with $p-1=s=6, 12$ in (\ref{mindims}). The chiral ring in this
minimal model case is generated by one generator $x$ obeying the
relation $x^{k+2}=x^{18}=0$. The chiral primaries present in the coset
correspond to the elements $x^6$ and $x^{12}$ which form a consistent
subring of the original chiral ring. The large $N$ structure of the
chiral ring can, we believe, be exploited to study the large $N$
physical characteristics of the gauged fermion model.

Finally, it would be very interesting to uncover the AdS dual to this
interesting class of CFTs (in the large $N$ limit). The simpler coset
models in (\ref{wnmin}) have been identified \cite{gabgop, gabgop2}
with a class of higher spin Vasiliev theories on $AdS_3$
\cite{vasmat}. These have a single bulk gauge field of every spin
$s\geq 2$. Roughly speaking there is a single Regge trajectory in the
bulk and no Hagedorn growth of states. This is related to the fact,
mentioned above, that there is a single conserved current in these
coset CFTs for a given spin. The $W$-symmetry in the coset considered
here is very different having many more fields. In general, we expect
there to be a Hagedorn density of states in this theory corresponding
to single trace words built from $\psi_L, \psi^\dagger_L, \psi_R,
\psi^\dagger_R$ together with derivatives (modulo the projection by
the diagonal $SU(N)$ currents). Thus the bulk dual is presumably a
string theory on $AdS_3$. This fits with the fact that these cosets
have a central charge proportional to $N^2$ as opposed to the cosets
in (\ref{wnmin}) which have $c \propto N$ and behave like vector
models with fewer gauge invariant states
\cite{Klebanov:2002ja}\cite{gabgop}.

Though the coset theory is a strongly interacting CFT as evidenced by
the anomalous dimensions of different operators, it is interesting
that the dimensions of the operators in the table \ref{tab:primaries}
are like those of a free theory in the large $N$ limit. They are
integer multiples of $1\over 3$. In that sense the spectrum is similar
to that of free Yang-Mills theory or that of the D1-D5 CFT
\cite{Dijkgraaf:1998gf, Seiberg:1999xz} at a symmetric orbifold
point. Note that these are the points where we expect a dual
tensionless string theory with an unbroken higher spin symmetry. This
is consistent with the fact that these cosets have a large chiral
$W$-algebra as discussed above.

\section{Low Density}
\label{sec:low}

In this section, we will study the behavior of the $SU(N)$ gauge
theory coupled to two adjoint multiplets of Majorana fermions for small $U(1)$
chemical potential $\mu$. We will work in the regime where $m$, the
mass of the adjoint, is much smaller than the scale set by the 't
Hooft coupling $g_{YM}\sqrt{N}$. We are interested in the low energy
dynamics of the system as $\mu$ is increased from zero to values
comparable to the scale set by $m$. Some features of this low energy
theory can be inferred from the spectrum of gauge invariant hadronic
states at $\mu=0$. Let us describe how this works for our model.

Generally, it is extremely difficult to compute the spectrum of
hadronic bound states in gauge field theories such as QCD. In $1+1$
dimensions, however, the light cone quantization can make this problem
tractable. Compactification of the light-like coordinate on a circle:
$x^-\sim x^- + L$, a formal regularization procedure known as the discrete light cone
quantization (DLCQ) \cite{Pauli:1985pv,Hornbostel:1988fb}, typically
reduces the problem to matrix diagonalization.  The problem is
further simplified in the planar (large $N$) limit where the $SU(N)$
singlet states
are non-interacting.  The physical bound state spectrum can be
inferred by taking the decompactification limit for the
light-cone coordinate $x^-$. This can be computationally expensive for certain
models, but is nonetheless a well controlled approximation
scheme. Often, this continuum limit is presented by taking the limit
$K\rightarrow \infty$, where the integer `harmonic resolution parameter' $K$ enters the definition
light cone momentum:
\begin{equation} P^+ = {\pi K \over L} \end{equation}
After diagonalization of the light cone hamiltonian, $P^-$,
the spectrum of bound state masses $M$ is read off from
\begin{equation} M^2 = 2 P^+ P^- \ .\end{equation}

In this paper we will be primarily concerned with the spectrum of
bound state masses in the regime where $m^2$ is kept fixed while the
`t Hooft coupling $g_{YM}^2 N$ is sent to infinity. We find a rich
spectrum of bound states whose masses divided by $m$ approach
constants in this limit. These bound states originate from the large
$N$ coset CFT discussed in section II, perturbed by the operator
$m(\Tr (\psi_L \psi_R^\dagger) + \rm {h.c.})$ of dimension $1/3$. This
mass operator breaks the $Q_L-Q_R$ symmetry of the CFT, but the
overall $U(1)$ charge symmetry, $Q=Q_L+Q_R$, remains
unbroken. Therefore, the bound states break up into sectors labeled by
the integer charge $Q$.

The DLCQ of a closely related system consisting of a single
adjoint multiplet of Majorana fermions has been analyzed in the past
\cite{igor1,igor2,igor3}.  However, in the same limit of sending the
't Hooft coupling to infinity while keeping $m^2$ fixed, all bound
states acquire masses of order $\sim g_{YM} \sqrt N$. The fact that
there are no light bound states with masses of order $m$ is due to the
triviality of the CFT arising from the $SU(N)$ gauge theory coupled to
an adjoint Majorana fermion: in that case, instead of
(\ref{centralcharge}) one finds $c=0$ \cite{igor2}.

We will summarize the details of the DLCQ computation in Appendix
C. In this section, we will focus on presenting the results of the
computation.

First, let us describe the basic physics of the model. In $1+1$
dimensions, gauge fields are non-dynamical but serve as agents binding
the colored matter fields. The hadronic bound states will then be
superpositions of traces of products of the adjoint creation operators
(\ref{states}). The problem can be separated into boson and fermion
sectors depending on whether the number of the creation operators in
the trace is odd or even.\footnote{The CFT arising in the $m=0$ limit
exhibits the ${\cal N}=2$ supersymmetry relating the bosonic and the
fermionic operators. Since the light cone quantization is unreliable
in the presence of massless states, we will keep $m$
non-vanishing. Then the supersymmetry is broken; so, the boson and
fermion bound state spectra are not identical. However, the bound
states with $m\ll M \ll g_{YM}\sqrt N$ may exhibit approximate
supersymmetry. Such highly excited states are difficult to access
numerically, but it would be very interesting to look for the emergent
supersymmetry in this region of bound state masses.}

When there are two adjoint Majorana fermions, there is a $U(1)=SO(2)$
global flavor symmetry which provides an additional quantum number to
label the states in the spectrum. One way to make this manifest is to
combine the two adjoint Majorana fermions into an adjoint Dirac
fermion, and count the difference between the number of fermions and
anti-fermions for each state.

Suppose for $\mu=0$ we succeed in computing the masses $M$ and charges
$Q$ of all the bound states. Suppose also that all states have $M>0$
and as a result the theory is gapped in the far IR. What happens when
$\mu$ is increased?

We expect that some of the particles will condense when
\begin{equation}\zeta = M - \mu Q \end{equation}
becomes negative.

If we were working in dimensions greater than $1+1$, then depending on
whether the first state for which $\zeta$ becomes negative is a boson
or a fermion, the system would exhibit the universal behavior of
either the Bose-Einstein condensation or formation of a degenerate
Fermi gas. In $1+1$ dimensions, the distinction is somewhat blurred by
the fact that bosons and fermions are related by bosonization. If the
spectrum is sufficiently generic so that there is precisely one state
for which $\zeta$ is going to zero at the minimal critical $\mu$, then
one expects the system to behave as a Luttinger liquid.

Additional subtleties can arise from the fact that there might be a
degeneracy causing more than one state to hit $\zeta=0$ at the same
time. Logically, there are three distinct possibilities.
\begin{enumerate}
\item[{\bf I}] The value of $\zeta$ goes to zero for exactly one state.
\item[{\bf II}] The value of $\zeta$ goes to zero for several, but a
finite number of states.
\item[{\bf III}] The value of $\zeta$ goes to zero for an infinite set of
degenerate states.
\end{enumerate}
The DLCQ computation of the bound state spectrum can help distinguish
among these three possibilities.

\begin{figure}
\centerline{\includegraphics{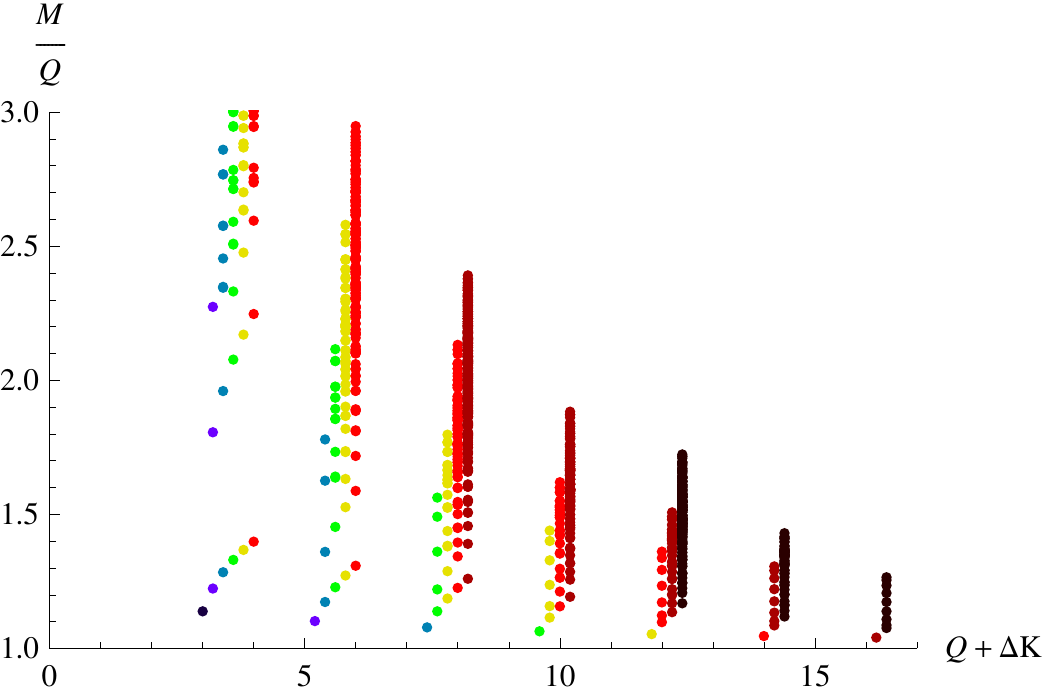}}
\caption{The spectrum of fermionic bound states for
$K=5,7,9,11,13,15,17,19$. The states with the same $K$ are shown using
the same color.  Increasing $K$ for fixed $Q$ is illustrated by a
gradual shift to the right in each of the columns. \label{fullspec}}
\end{figure}

The details regarding the implementation of the DLCQ procedure are
summarized in the appendix. Here we will merely present the result,
where we display the full spectrum in figure \ref{fullspec}.

The points illustrated in figure \ref{fullspec} are to be interpreted
as follows.
\begin{enumerate}
\item These points correspond to the spectrum of fermions for
which the harmonic resolution parameter $K$ takes odd integer values. One
expects to recover the exact spectrum in the limit $K \rightarrow
\infty$.
\item The `t Hooft coupling $g_{YM}^2 N$ is taken, for the sake of
definitiveness, to be $2\pi \cdot 10^3$ times the bare mass-squared of
the fermions, $m^2$. As long as this number is very large, the spectrum in
the range illustrated, presented in units where $m^2=1$, is
insensitive to its precise value. The idea is to extract the behavior
of this system in the limit of large $g_{YM}^2 N$.
\item $M$ is the mass of the hadronic bound state. Here we are
displaying $M/Q$. The state with lowest $M/Q$ is the one whose $\zeta$
will hit zero first as $\mu$ is increased.
\item For each $K$, $Q$ in the range from $Q=1$ to $Q=K$ are
possible. However, the $Q=1$ states are heavier than the range of $M$
illustrated in figure \ref{fullspec} and as a result are not displayed
in this figure. The state in the $Q=K$ sector is decoupled from the rest
of the dynamics. In general, taking the large $K$ limit with $Q$ fixed
will give rise to a reliable extrapolation of the spectrum of that
fixed $Q$ sector. The spectrum with $Q$ of the order of $K$, however,
is sensitive to the DLCQ artifacts and does not effectively approximate
the spectrum in the continuum limit. Here we have computed and
presented the states for $Q$ in the range $3 \le Q \le K-2$ with the
exception of some small $Q$ states for large values of $K$ for which
the computations became numerically intractable.
\item For each value of $K$, states with different charges $Q$
are displayed in separate columns. States with the same
$K$, however, can be identified by the fact that they are plotted
using the same color. For each $Q$, increase in $K$ is indicated by
gradual shift in the column of points to the right.

\end{enumerate}

In order to identify the states with smallest $M/Q$, one must, for
each $Q$, track the lowest mass state and extrapolate to large values
of $K$.  The spectrum illustrated in figure \ref{fullspec} suggests
that, for each $Q$, the masses are gradually increasing in a similar
manner as $K$ is increased. It also suggests that these increasing
masses are converging to the large $K$ limit.

\begin{figure}
\centerline{\includegraphics{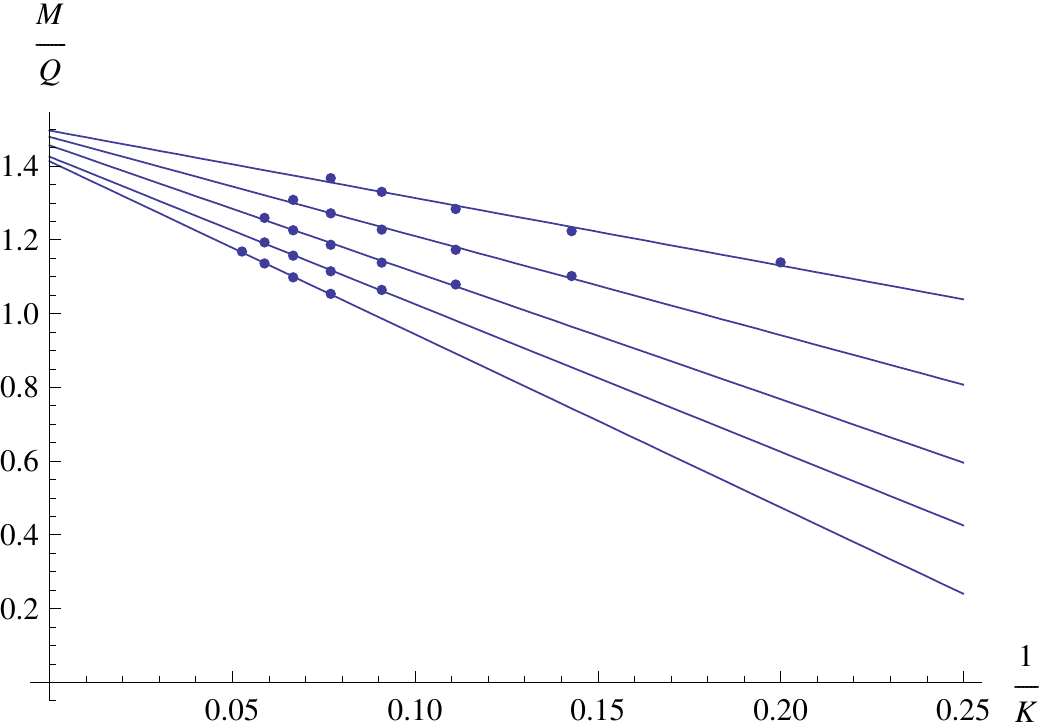}}
\caption{The spectrum $M/Q$ of lightest states as a function of $1/K$
for $Q=3,5,7,9,11$. The lines are linear fits to the available
data. The points for $Q=3$ have the largest $M/Q$. For each successive
$Q$, the $M/Q$ is decreasing. The intercept of the linear fit at
$1/K=0$ is the extrapolated value of the $M/Q$ for the lightest state
for fixed value of $Q$.
\label{extrapolate}}
\end{figure}

In order to analyze the asymptotic behavior of the masses of the
low-lying states, it is useful to plot their masses as a function of
$1/K$. Since we are interested in how these states are affected by the
chemical potential, we will actually plot $M/Q$ as a function of $1/K$
for different values of $K$. This is illustrated in figure
\ref{extrapolate} for $Q=3,5,7,9,11$.

We have also superposed a line indicating the linear extrapolation
available from the set of data available. These lines cross $1/K=0$ at
finite values. This can be viewed as a crude method to extract the
extrapolated value for the large $K$ limit.

What we see in figure \ref{extrapolate} is the tendency for the large
$K$ limit of the $M/Q$ of the lightest states to become dense with
increasing $Q$.  Since $Q$ can get arbitrarily large, this suggests
that in the large $K$ limit, states with arbitrarily large $Q$ are
converging to the same value of $M/Q$.  In other words, we seem to be
finding out that our model is exhibiting scenario {\bf III} enumerated
earlier in this section.  It should be kept in mind, however, that one
must send $K \rightarrow \infty$ first, and then look for a trend as
$Q$ is increased. The order of these two limits cannot be exchanged.

It is not easy to determine the critical value $\mu_{crit}$ of $M/Q$.
To settle this question, a higher precision computation is
required. Unfortunately, for the reason given above, one must explore
very large values of $K$ in order to explore large values of $Q$. This
is computationally very expensive.

The fact that there may be infinitely many states with degenerate
$\mu_{crit}=M/Q$ suggests that, when $\mu$ approaches $\mu_{crit}$,
the system may undergo a transition from a gapped phase into a phase
with a non-trivial interacting conformal field theory with $c>1$ in
the IR.

The DLCQ results presented in this section focused mainly on fermionic
bound states. We have also carried out the analogous computation for
bosonic bound states and found a similar behavior of $M/Q$. We will
not present the detailed results of our analysis for the bosonic
spectrum here. Some technical issues which arise for the DLCQ
computation in the bosonic sector will be described briefly in
Appendix C.

Let us, in closing, mention that since all of these results were
presented for the case where $N$ was taken to infinity first, it is
certainly possible for the seemingly dense spectrum of $M/Q$ to be
discretized by a fine structure of order $N^{-\alpha}$ for some
$\alpha>0$. Such a structure can change the basic feature of our model
from scenario {\bf III} to {\bf II} or {\bf I}. This may also be
related to the fate, following the relevant flow, e.g.\
(\ref{eq:mostrelevant}), of the coset CFT discussed in the earlier
sections.

\section{Conclusions}
\label{sec:conc}

In this paper, we described a novel metallic state of matter in one
spatial dimension, with a continuously variable density.

The well-studied one-dimensional metallic state is the Luttinger
liquid, and in many respects this state can be considered to be the
natural limiting case of the Fermi liquid state of higher dimensions.
Indeed, the Luttinger liquid reduces to a free fermion model at a
specific value of the Luttinger parameter, and other parameter values
are continuously connected to this one. It has central charge $c=1$,
and this is directly linked to the massless scalar representing
fluctuations of the globally conserved $U(1)$ charge density.
We note that some Luttinger liquids have additional gapless modes
associated with other global symmetries (and so a larger central charge),
such as the models described in Appendix~\ref{app:fund}.

Our state was obtained by considering a non-zero density of Dirac
fermions carrying {\em adjoint\/} color charges of a $SU(N)$ gauge
field. We found a `strange metal' state described by two-dimensional
superconformal field theories with central charges $c=(N^2-1)/3$. For
large $N$ the central charge becomes large, while the global symmetry
remains only the $U(1)$ associated with fermion number
conservation. The strange metal has a Fermi surface with a Fermi
wavevector, given by (\ref{fermi}), which is equal to that of
non-interacting color-charged particles; the Fermi wavevector changes
continuously as the density is varied.  This Fermi surface is `hidden'
\cite{hyper}, because the single fermion Green's function is not a
gauge-invariant observable.  Nevertheless, the Fermi surface and the
value of the Fermi wavevector are detectable in the Friedel
oscillations of (\ref{friedel}).  We propose that this variable
density state, with its large phase space of low energy excitations
linked to its large central charge and Fermi surface of color-charged
fermions, can serve as a paradigm for non-Fermi liquid states in two
and higher spatial dimensions.

The structure of our $d=1$ strange metal is quite analogous to the
`hidden' Fermi surface states obtained recently for general $d$ in
Refs.~\cite{liza,tadashi1,hyper,evajoe} via the AdS/CFT
correspondence.  In Refs.~\cite{tadashi1,hyper} it was postulated that
the Fermi surfaces of gauge-charged particles could be detected by the
hyperscaling violation of the thermal entropy density, and by a
logarithmic violation of the `boundary law' of entanglement entropy.
The coefficient of the entanglement entropy logarithm was used to
deduce a value for the Fermi wavevector which depended upon
ultraviolet details only through the value of the density $Q/L^d$, in
just the manner expected from the Luttinger relation \cite{hyper}.  In
our $d=1$ strange metal here, analogous properties of the entropy and
entanglement entropy are trivially satisfied, because the hyperscaling
violation exponent $\theta = d-1=0$ (in the notation of
Ref.~\cite{hyper}), and every CFT$_2$ has a logarithmic violation of
the boundary law of entanglement entropy \cite{holzhey,cardycal}.
However, here we were able to detect the Fermi surface, and determine
the Fermi wavevector, from the Friedel oscillations in the density
correlations. Obtaining the Friedel oscillations in the AdS/CFT
description of strange metals in general $d$ is clearly one of the
important challenges for future work.  In this direction, it would be
useful to understand the large $N$ dependence of the proportionality
constant in (\ref{friedel}): this should shed light on how Friedel
oscillations of gauge-charged particles appear in holographic
theories. We also note the interesting recent computation of
\cite{evajoe} which detected Friedel oscillations in an anisotropic
quantum liquid of strings in $5+1$ dimensions.

In section III we studied the bound state spectrum of the large $N$ gauge
theory in the limit $g_{YM}^2 N \gg m^2$ using the numerical DLCQ
approximation. This approach sheds light on the properties of the
$SU(N)$ gauge theory at low density.  A very interesting phenomenon
that we have uncovered is the emergent ${\cal N}=2$ supersymmetry of
the gauge theory in the limit $m\rightarrow 0$. A useful direction for
future work would be to obtain some numerical DLCQ evidence for the
emergent supersymmetry by studying the masses of highly excited bound
states with $g_{YM}\sqrt N \gg M\gg m$. Hopefully, this spectrum will exhibit
approximate supersymmetry.

Another intriguing direction for future work, which was discussed at
the end of section II.C, is the possibility of a dual description of
the large $N$ CFT in terms of a theory with higher-spin gauge symmetry in
$AdS_3$. The existence of such a dual description is suggested by the
large $W$-symmetry, and by the fact all the operator dimensions appear
to approach constants in the large $N$ limit that are quantized in
units of $1/3$.  A useful $3+1$ dimensional analogue of the ${\cal
N}=2$ supersymmetric large $N$ CFT we are considering may be the
${\cal N}=4$ supersymmetric $SU(N)$ gauge theory, which is dual to the
$AdS_5\times S^5$ background of type IIB string theory
\cite{Maldacena:1997re,Gubser:1998bc,Witten:1998qj}. Both theories
have anomaly coefficients of order $N^2$. When the ${\cal N}=4$ gauge
theory has a large `t Hooft coupling $g_{YM}^2 N$, the dual
$AdS_5\times S^5$ background becomes weakly curved. In this limit the
dimensions of all operators not protected by supersymmetry become very
large. On the other hand, at vanishing `t Hooft coupling all the
operators have integer dimensions, and their number exhibits the
exponential Hagedorn growth. In this limit the curvature of the dual
string background becomes very large; this is often referred to as the
``tensionless string limit.'' It has been suggested
\cite{Sundborg:2000wp} that a useful dual description of the free
${\cal N}=4$ SYM theory may involve higher-spin gauge theory in
$AdS_5$ \cite{Vasiliev:2003ev} coupled to an infinite number of
additional fields.

Similarly, if an $AdS_3$ string dual of our ${\cal N}=2$
supersymmetric coset CFT is found, we expect it to be strongly curved.
It is therefore interesting to ask if the coset CFT has an exactly
marginal operator which could correspond to increasing the radius of
the dual background. In fact, the CFT has an exactly marginal
double-trace operator which is a product of the left and right $U(1)$
currents, $J_L(z) J_R(\bar z)$.  This operator breaks the ${\cal N}=2$
supersymmetry as well as some of the the extended $W$-symmetries. For
the $N=2$ coset CFT (\ref{entwocoset}), this marginal operator changes
the radius of the compact scalar in the bosonized formulation. In the
limit of large radius, a large gap develops between the dimensions of
typical momentum and winding operators. It would be interesting to
study the effect of the marginal deformation on the operator
dimensions for $N>2$, and to see if deforming the CFT along this
marginal direction could also create a large gap in the spectrum of
operator dimensions. The presence of such a gap would suggest
\cite{Heemskerk:2009pn} a weakly curved $AdS_3$ dual of the large $N$
CFT.

\acknowledgements

We thank J.~de~Boer, P.~Fendley, M.~R.~Gaberdiel, R.~Nandkishore,
E.~Witten, and J.~Zaanen for helpful discussions.  We also acknowledge
the provision of computing resources by the Condor Project at the
University of Wisconsin-Madison.  RG's work is supported in part by a
Swarnajayanthi fellowship of the DST and more broadly by the generous
funding for basic sciences by the people of India.  He would like to
thank ESI, Vienna, Newton Institute, Cambridge and ICTS, Bangalore for
hospitality during various stages of this work.  The work of AH was
supported by the DOE grant DE-FG02-95ER40896.  The work of IRK was
supported in part by the US NSF under Grant No.~PHY-0756966. IRK
gratefully acknowledges support from the IBM Einstein Fellowship at
the Institute for Advanced Study, and from the John Simon Guggenheim
Memorial Fellowship.  SS was supported by the National Science
Foundation under grant DMR-1103860, by a MURI grant from AFOSR.  KjS
thanks the Galileo Galilei Institute for Theoretical Physics in
Florence for the hospitality and the INFN for partial support during
the completion of this work.

\appendix

\section{Fundamental matter}
\label{app:fund}

This appendix briefly describes the high density physics of the theory
in (\ref{L}), but for the case where $\Psi$ transforms as a
fundamental of the $SU(N)$ gauge group.  For completeness, we
include the case where
$\Psi$ has a flavor index which takes $N_f$ values, and then the model
has a $U(N_f)$ global symmetry. It is convenient to decompose the
global symmetries into a $U(1)$ symmetry associated with the `charge'
density, and a $SU(N_f)$ flavor symmetry (the latter is absent for
$N_f=1$).  We proceed just as in section~\ref{sec:high}. The high
density limit is characterized by a Fermi wavevector of gauge-charged
fermions given by
\beq
\frac{Q}{L} = N N_f \frac{k_F}{\pi};
\eeq
note that the r.h.s.\ has a prefactor of $N$, rather than the $(N^2 -
1)$ in (\ref{fermi}).  The fluctuations near this wavevector map onto
the $m=\mu=0$ theory, which was considered in
\cite{gonzalez,affleck1}.  Now we bosonize the $N N_f$ complex Dirac
fermions differently. Rather than considering them as $2N N_f$
Majorana fermions, we note that they can be used generate WZW currents
of $SU(N)_{N_f}$, $SU(N_f)_{N}$, and $U(1)$, and these fully span the
Hilbert space \cite{gonzalez,affleck1,frishman}.  The $SU(N)_{N_f}$
currents are projected out by the gauge field, and so the low energy
theory is made up of two decoupled sectors: a $c=1$ free gapless
scalar associated with the $U(1)$, and a $SU(N_f)_N$ WZW model
associated with the $SU(N_f)$ global flavor symmetry with
$c=N(N_f^2-1)/(N+N_f)$. This decoupling of the $U(1)$ density mode is
the key simplifying feature of the fundamental matter case, and was
absent for the adjoint matter case considered in the body of the
paper.  Consequently, the $U(1)$ sector here is similar to an ordinary
Luttinger liquid, while the spectator $SU(N_f)_N$ WZW model is
directly linked to the additional global flavor symmetries of the
model.

In the notation of \cite{qpt2},
we can write the Hamiltonian of the $U(1)$ sector as
\beq
H = \frac{1}{2 \pi} \int dx \left[ \frac{1}{K} \left( \partial_x \phi \right)^2
+ K  \left( \partial_x \theta \right)^2 \right]
\eeq
where $K$ is the Luttinger parameter, $\theta$ and $\phi$ are scalar fields
obeying the commutation relations
\beq
[\partial_x \phi (x), \theta(x')] = [\partial_x \theta (x), \phi (x')] = i \pi \delta (x-x'),
\eeq
and the $U(1)$ charge is
\beq
Q = \frac{1}{\pi} \int dx \, \partial_x \phi.
\eeq
The variable $K$ is related to the exactly marginal perturbation to
the $U(1)$ theory, the analog of the `radius' of the scalar in the
string theory notation.  The fermion fields are related to these
scalar fields via \cite{afflud}
\beq
\psi_{R,L} = e^{- i (\theta \pm \phi)/\sqrt{N N_f}} \, \varphi_{R,L}^c \, \varphi_{R,L}^f \label{fermibose}
\eeq
where $\varphi_{R,L}^c$ is the primary field the $SU(N)_{N_f}$ WZW
model transforming as a fundamental of $SU(N)$, $\varphi_{R,L}^f$ is
the primary field the $SU(N_f)_{N}$ WZW model transforming as a
fundamental of $SU(N_f)$. The exponential factor has been chosen so
that the free fermion theory without the $SU(N)$ gauge field is
properly bosonized at $K=1$ with both WZW models conformal so that
$\mbox{dim}[ \varphi_{R,L}^c ] = (N^2 -1 )/(2 N (N+N_f))$ and
$\mbox{dim}[ \varphi_{R,L}^f ] = (N_f^2 -1 )/(2 N_f (N+N_f))$.

To determine the Friedel exponent, we need the smallest scaling
dimension operator with $Q_L=1$ and $Q_R=-1$. Applying
(\ref{fermibose}) to the operator $\mbox{Tr} (\psi_L^\dagger \psi_R)
$, we can set the trace over the $SU(N)_{N_f}$ WZW fields to constants
\cite{affleck1,frishman}, and the scaling dimensions of the remaining
sectors yield
\beq
\Delta_F = \mbox{dim}[ e^{2i\phi/\sqrt{N N_f}}] + 2 \, \mbox{dim}[\varphi_L^f] = \frac{K}{N N_f} + \frac{(N_f^2 - 1)}{N_f (N+N_f)},
\eeq
where $K$ is now allowed to be not equal to unity because, in general,
there will be an exactly marginal interaction in the $c=1$ sector.  We
note that the $\Delta_F \rightarrow 0$ corresponds to a crystalline
state with broken translational symmetry; such continuous symmetry
breaking is not possible in 1+1 dimensions, but a crystal was
discussed as a mean field theory of baryons valid in the formal large
$N$ limit \cite{schon}.

In the fundamental matter model, the pairing operator $ \psi_L \psi_R$
can be reduced to a gauge singlet only for $N=2$. Extending our
Friedel operator argument to a gauge singlet pairing operator implies
that we need the simplest operator with $Q_L=Q_R=-1$, and this yields
\beq
\Delta_P = \mbox{dim}[ e^{2i\theta/\sqrt{N N_f}}] + 2 \, \mbox{dim}[\varphi_L^f] = \frac{1}{N N_f K}+ \frac{(N_f^2 - 1)}{N_f (N+N_f)}
\quad , \quad N=2.
\eeq

Finally, we note that the case $N=1$ and $N_f=1$ corresponds to the finite density phase of the massive Thirring model,
which realizes the simplest Luttinger liquid.

\section{Modular invariants for coset CFT$_2$}
\label{app:modinv}

Modular invariant partition functions for coset CFT$_2$ can be
constructed once invariants for both the numerator and denominator
CFT$_2$'s have been specified \cite{bouwknegt,bouwknegt2}.  Rather
than describing the general construction, we focus on the specific
example of the $(N,N;2N)$ cosets for gauged adjoint fermions.

The numerator CFT$_2$ has two copies of the $SU(N)_N$ theory, each
describing $(N^2-1)$ adjoint fermions. There are several options for a
modular invariant, including a simple product of the diagonal modular
invariant of each of the $SU(N)_N$ factors. However, we should here
select the invariant that describes the situation where boundary
conditions on the combined fermions are such that a global $U(1)$
symmetry arises. The appropriate modular invariant turns out to be the
diagonal modular invariant of an $SO(2N^2-2)_1$ symmetry, written as
\beq
Z^{SO(2N^2-2)_1} = |\chi_1^{SO(2N^2-2)_1}|^2 + |\chi_{\rm v}^{SO(2N^2-2)_1}|^2+ 2|\chi_{\rm sp}^{SO(2N^2-2)_1}|^2
\eeq
with `$1$', `v,' and `sp' denoting the identity, vector and spinor
representations of $SO(2N^2-2)_1$.  This result arises from the well
known result, known as non-Abelian bosonization \cite{witten}, that
the CFT$_2$ based on $SO(M)_1$, at central charge $c=M/2$, describes
$M$ real fermions.

The partition sum can be re-expressed in terms of characters of two
copies of $SO(N^2-1)_1$, one for each of the groups of $N^2-1$
fermions
\begin{eqnarray}
\lefteqn{Z^{SO(2N^2-2)_1} =
|\chi_1^{SO(N^2-1)_1} \widetilde{\chi}_1^{SO(N^2-1)_1}+ \chi_{\rm v}^{SO(N^2-1)_1} \widetilde{\chi}_{\rm v}^{SO(N^2-1)_1} |^2}
\nonumber \\[2mm]
&&
+ | \chi_{\rm v}^{SO(N^2-1_1)} \widetilde{\chi}_{\rm 1}^{SO(N^2-1)_1} + \chi_{\rm 1}^{SO(N^2-1)_1} \widetilde{\chi}_{\rm v}^{SO(N^2-1)_1}|^2
+ 2\lambda  |\chi_{\rm sp}^{SO(N^2-1)_1} \widetilde{\chi}_{\rm sp}^{SO(N^2-1)_1}|^2 \ ,
\end{eqnarray}
with $\lambda=4(1)$ for $N$ odd(even).The $SO(N^2-1)_1$ characters can
each be branched into characters of $SU(N)_N$. For the NS sector
characters (labels `1' and `v') the results are
\bea
&&  \chi^{SO(N^2-1)_1}_1 =
\chi^{SU(N)_N}_{(00\ldots00)} + \chi^{SU(N)_N}_{(20\ldots 10)} + \chi^{SU(N)_N}_{(01\ldots02)} + \ldots
\nonumber\\[2mm]
&&  \chi^{SO(N^2-1)_1}_{\rm v} =
\chi^{SU(N)_N}_{(10\ldots01)} + \chi^{SU(N)_N}_{(110\ldots011)} + \dots
\eea
We use Dynkin labels $(l_1l_2\ldots l_{N-1})$ to tag the $SU(N)$
representions: $(00\ldots00)$ is the identity, $(10\ldots01)$ the
adjoint, etc. (a useful reference for the group theory is
\cite{slansky}).  We remark that the $SU(N)$ representations featuring
in the NS sector satisfy the $N$-ality condition
\beq
l_1 + 2 l_2 + \ldots (N-1) l_{N-1} \equiv 0 \mod N \ .
\label{eq:Nality}
\eeq

To obtain a partition sum for the coset CFT$_2$ the following steps
are taken. First one writes branching rules for the $SO(2N^2-2)_1$
characters into products of branching functions times characters of
the affine algebra $SU(N)_{2N}$ that features in the denominator of
the coset. Schematically
\beq
\chi^{SO(2N^2-2)_1}_A = \sum_a   b_A^a \times \chi^{SU(N)_{2N}}_a \ .
\label{eq:branching}
\eeq
The labels $a$ take values in the integral dominant weights of
$SU(N)_{2N}$, which are written as Dynkin labels $(l_1 l_2 \ldots
l_{N-1})$ satisfying $\sum_il_i\leq 2N$. We find that, for general
$N$, the terms on the r.h.s.\ of (\ref{eq:branching}) are grouped into
combinations of the form
\beq
\chi^{SU(N)_{2N}}_a + \chi^{SU(N)_{2N}}_{\lambda(a)} + \chi^{SU(N)_{2N}}_{\lambda^2(a)} + \ldots
\eeq
where $\lambda$ is the automorphism
\beq
\lambda:   ( l_1 l_2 \ldots l_{N-1}) \to ([2N-\sum_i l_i] l_1 l_2 \ldots l_{N-2}) \ .
\label{eq:auto}
\eeq
Again, the $SU(N)$ representations featuring in the NS sector satisfy
the $N$-ality condition (\ref{eq:Nality}).

Writing the modular invariants for the denominator (d) and numerator (n) as
\bea
&&
Z^{\rm d} = \sum_{AB} N^{\rm d}_{AB} \chi^{SO(2N^2-2)_1}_A \overline \chi^{SO(2N^2-2)_1}_B \ ,
\nonumber \\
&&
Z^{\rm n} = \sum_{ab} N^{\rm n}_{ab}  \chi^{SU(N)_{2N}}_a \overline{\chi}^{SU(N)_{2N}}_b
\eea
the coset invariant is obtained as
\beq
Z^{\rm coset} = \sum_{ABab} N^{\rm d}_{AB} N^{\rm n}_{ab} b_A^a \overline{b}_B^b  \ .
\eeq

For the $(N,N;2N)$ cosets, the natural choice for the denominator
modular invariant is the $SU(N)_{2N}$ invariant that displays the same
grouping of characters, according to the automophism (\ref{eq:auto}),
that we observed in the branching rules (\ref{eq:branching}).  Such an
invariant exists for general $N$ \cite{gannon2}; we display explicit
examples for $N=2,3$ in the main text of this paper.

The ${\cal N}=(2,2)$ superconformal symmetry of the $(N,N;2N)$ coset
guarantees that the branching functions $b^a_A$ are characters of (an
extension) of the ${\cal N}=2$ superconformal algebra.  For $N\geq 3$
we find (see again the main text) that the vacuum character of this
extended symmetry takes the form
\beq
{\rm ch}^{{\cal N}=2,{\rm ext}}_{q=0,h=0}
=
{\rm ch}^{\rm NS}_{q=0,h=0}+  {\rm ch}^{\rm NS}_{q=1/3,h=2}+  {\rm ch}^{\rm NS}_{q=-1/3,h=2}+ \ldots
\eeq
The $(N,N;2N)$ coset modular invariant can be viewed as a diagonal
modular invariant for this $W$-extension of ${\cal N}=2$
superconformal symmetry.

\section{DLCQ Quantization of the gauged adjoint Dirac fermions in 1+1 dimensions}
\label{app:DLCQ}

In this appendix, we will summarize the computation of the discretized
light cone quantization spectrum of the $1+1$ dimensional $SU(N)$
gauge theory coupled to adjoint Dirac fermions in the large $N$ limit.

The first step is to write down the Lagrangian which follows the
general pattern of \cite{igor1,igor2,igor3} except that the fermions
are now complex. Here we follow II.B of \cite{igor1}. We start with
(1) of \cite{igor3} but treat
\begin{equation} \Psi = {2^{1/4}} \left(\begin{array}{c} \psi \\ \chi \end{array}\right)
\end{equation}
as Dirac fermions.

The light cone coordinates are defined by
\begin{equation} x^\pm = {1 \over \sqrt{2}} (x^0 \pm x^1) \end{equation}
so that
\begin{equation} A_\pm = {1 \over \sqrt{2}} (A_0 \pm A_1) \ . \end{equation}

We will use for the Dirac matrices
\begin{equation} \gamma^0 = i \sigma_2 = \left(\begin{array}{cc} 0 & -i \\ i & 0 \end{array}\right),
\qquad
 \gamma^1 = i \sigma_1 = \left(\begin{array}{cc} 0 & i \\ i & 0 \end{array} \right) \ . \end{equation}

The Lagrangian is normalized to take the form
\begin{equation}  \mbox{Tr} \left\{\bar{\Psi} ( i \partial \!\!\!/ - m) \Psi]\right\}   = \mbox{Tr} \left\{2i \psi^\dag \partial_+ \psi  +  2 i \chi^\dag \partial_- \chi - i  \sqrt{2} m (\psi^\dag \chi + \chi^\dag \psi) \right\}. \label{app31} \end{equation}
To compare this Lagrangian with the Hamiltonian (\ref{h0}), simply
note that the fermion field $\Psi$ can be expressed in the standard
mode expansion
\beq 
\label{modexpan}
\Psi(t,x) = \int {d k_1 \over 2 \pi} {1 \over \sqrt{2 k_0}}\left( u(k) p(k) e^{-i k_\mu x^\mu}  + v(k) h(k) e^{i k_\mu x^\mu}\rule{0ex}{3ex}\right) \eeq
where $u(k)$ and $v(k)$ is the standard 2 component spinor basis satisfying
\beq (k\!\!\!/ - m ) u(k)=  (k\!\!\!/ + m ) v(k)=  0, \qquad \bar u(k) u(k) = - \bar v(k) v(k) = 2m \ . \label{diracspinor} \eeq
Then, in terms of $p(k)$ and $h(k)$ we recover (\ref{h0}) for the
Hamiltonian.

Returning to (\ref{app31}), we gauge the free fermion theory by
introducing covariant derivatives
\begin{equation} D\Psi  = \partial_\mu \Psi  + i [A_\mu,\Psi] \ . \end{equation}

It is customary in light cone quantization to use the gauge
\begin{equation} A_-=0 \end{equation}
so that the gauge kinetic term takes the form
\begin{equation} -{1 \over 2 g_{YM}^2} {\mbox Tr} F^2 = {1 \over g_{YM}^2} \mbox{Tr} (\partial_- A_+)^2 \end{equation}
and the Lagrangian reads
\begin{equation} {\cal L} = \mbox{Tr} \left\{2i \psi^\dag \partial_+
\psi + 2 i \chi^\dag \partial_- \chi - \sqrt{2} i m
(\psi^\dag \chi + \chi^\dag \psi) + A_+ J^+ + {1 \over
{g_{YM}^2}} (\partial_- A_+)^2\right\} \end{equation}
with
\begin{equation} J^+ =  2 (\psi \psi^\dag + \psi^\dag \psi)\ . \end{equation}
If we take $x^+$ as the time direction, $\chi$ and $A_+$ are
non-dynamical and can be integrated out, giving rise to the light-cone
momentum and Hamiltonian
\begin{eqnarray}
P^+ & = & \int dx^- \, \mbox{Tr} \left\{ 2i \psi^\dag \partial_- \psi\right\}\label{Pp} \ , \\
P^- & = & \int dx^- \, \mbox{Tr} \left\{  -  i m^2  \psi^\dag {1 \over \partial_-} \psi  - {1 \over {4}} g_{YM}^2 J^+ {1 \over \partial_-^2} J^+ \right\} \ . \label{Pm}
\end{eqnarray}
Imposing canonical quantization on the fermions gives rise to relation
\begin{equation} \{ \psi^\dag_{ij}(x^-) , \psi_{kl}(\tilde x^-) \} = {1 \over 2} \delta(x^- - \tilde x^-) \left( \delta_{il} \delta_{jk}  - {1 \over N} \delta_{ij}\delta_{kl} \right) \label{cc2} \end{equation}
with
\begin{equation} \{\psi_{ij}(x^-), \psi_{kl}(\tilde x^-) \} = \{\psi_{ij}^\dag(x^-), \psi_{kl}^\dag(\tilde x^-) \} = 0 \ . \end{equation}
The Dirac fermions are expanded in modes
\begin{equation} \psi(x^-) = {1 \over \sqrt{2L} } \sum_{n \in {\rm odd} > 0} \left\{ A_{ij}(n) e^{-i \pi n x^-  / L}
+ B_{ji}^\dag (n) e^{i \pi n x^-  / L}\right\}  \end{equation}
where we have compactified the $x^-$ direction and imposed the
anti-periodic boundary condition on the $\psi(x^-)$ field; this
typically leads to a better DLCQ computation than choosing the
periodic boundary condition and removing the zero mode by hand. The
choice of boundary condition should not matter in the
decompactification limit $K\rightarrow \infty$.

The anti-commutation relation for the modes are is
\begin{eqnarray} \{A_{ij}(m),A_{kl}(n)\}  &=& \delta(m+n) \left(\delta_{il} \delta_{jk} - {1 \over N} \delta_{ij} \delta_{kl}\right) , \\
\{B_{ij}(m),B_{kl}(n)\}  &=& \delta(m+n) \left(\delta_{il} \delta_{jk} - {1 \over N} \delta_{ij} \delta_{kl} \right)
\end{eqnarray}
where $n$ takes odd integer values, and
\begin{equation} A_{ij}(-n) = A^\dag_{ji}(n), \qquad  B_{ij}(-n) = B^\dag_{ji}(n)  \ . \end{equation}
We can now set up the light cone vacuum
\begin{equation} A_{ij} (n) |0 \rangle = B_{ij} (n) |0 \rangle = 0, \qquad (n>0) \ . \end{equation}

The states are then generated by acting by a string of ``letters''
$A(-n)$ and $B(-n)$ in a single trace state, i.e.\
\begin{equation} | \psi \rangle  = \# \mbox{Tr} (B(-n_1) A(-n_2) ... B(-n_k)) | 0 \rangle \label{states}
\end{equation}
where $\#$ is a symmetry factor to ensure that the norm of each
state is one.

Our next step is to write the light cone momentum and Hamiltonian
operators in terms of the fermion oscillators. It is clear that the
light cone momentum operator (\ref{Pp}) can be written in the form
\begin{equation} P^+ =  \sum_{n \ge 1} \left\{    {\pi n \over L}  A^\dag_{ij}(n) A_{ji}(n) +      {\pi n \over L}  B^\dag_{ij}(n) B_{ji}(n)  \right\}= {\pi K \over L}  \ ,\end{equation}
when acting on a state with fixed $K$.

Instead of writing the light cone Hamiltonian $P^-$ in terms of
fermion oscillators, let us consider the Lorentz invariant mass
operator
\begin{equation} 2 P^+ P^- = V + T \end{equation}
where $V$ corresponds to terms associated with the term proportional
to $m^2$ and $T$ corresponds to the term proportional to $(J^+)^2$ in
(\ref{Pm}). Then, it is easy to show that
\begin{equation} V=
K m^2 \sum_{n \ge 1} \left\{
 \left({1 \over    n} \right) A_{ij}^\dag(n) A_{ji}(n)
+  \left({1 \over    n} \right) B_{ij}^\dag(n) B_{ji}(n) \right\} \ .
\end{equation}
Note that the dependence on $L$ drops out, but there is still a
dependence on $K$.

Computation of $T$ involves a somewhat tedious exercise of normal
ordering the $(J^+)^2$ written in terms of fermion oscillator
operators. One can organize $T$ in accordance to the number of
oscillators destroyed and created.
\begin{equation} T = T_{1 \rightarrow 1} + T_{1 \rightarrow 3} + T_{2 \rightarrow 2} + T_{3 \rightarrow 1} \ .
\end{equation}
In this form, one finds after some algebra, that
\begin{eqnarray}
T_{1 \rightarrow 1} &=& {{2} g_{YM}^2 N K \over \pi} \sum_n \sum_{m=1}^{n-2} \left\{\left({1 \over n-m}\right)^2 A_{ji}^\dag(n) A_{ji}(n)+
\left({1 \over n-m}\right)^2 B_{ji}^\dag(n) B_{ji}(n)\right\}\\
T_{1 \rightarrow 3} & = &
\left({g_{YM}^2 N K \over {2}\pi}\right) \sum \delta(n_1 + n_2 +n_3 - m_1) \times \cr
&& \left \{ {2 \over (n_1-m_1)^2}
B^\dag_{ik}(n_3) A^\dag_{kl}(n_2) A^\dag_{lj}(n_1) A_{ij} (m_1) \right.\cr
&& \left({2 \over (n_1-m_1)^2} -{2 \over (n_3 - m_1)^2} \right)
A^\dag_{ik}(n_3) B^\dag_{kl}(n_2)  A^\dag_{lj}(n_1) A_{ij} (m_1) \cr
&& -{2 \over (n_3-m_1)^2}
A^\dag_{kl}(n_3) A^\dag_{lj}(n_2) B^\dag_{ji}(n_1)  A_{ki}(m_1) \cr
&& {2 \over (n_1-m_1)^2}
A^\dag_{lj}(n_3) B^\dag_{ji}(n_2) B^\dag_{ik}(n_1) B_{lk} (m_1)  \cr
&& \left({2 \over (n_1-m_1)^2}-{2 \over (n_3-m_1)^2}\right)
B^\dag_{ji}(n_3) A^\dag_{ik}(n_2) B^\dag_{kl}(n_1) B_{jl}(m_1) \cr
&& \left. -{2 \over (n_3-m_1)^2}
B^\dag_{ji}(n_3) B^\dag_{ik}(n_2)  A^\dag_{kl}(n_1)  B_{jl}(m_1)\right\} \label{ham1} \\
T_{2 \rightarrow 2} & = &
\left({g_{YM}^2 N K \over {2}\pi}\right) \sum \delta(n_1 + n_2 - m_1 - m_2) \times \cr
&& \left\{ {2 \over (n_1-m_1)^2}
A^\dag_{kl}(n_2) A^\dag_{lj}(n_1) A_{ki}(m_2) A_{ij} (m_1)\right.\cr
&& \left({2 \over (n_1-m_1)^2}-{2 \over (m_1+m_2)^2}\right)
A^\dag_{ik}(n_2) B^\dag_{kl}(n_1) A_{ij}(m_2) B_{jl}(m_1)\cr
&& -{2 \over (m_1+m_2)^2}
B^\dag_{ik}(n_2)A^\dag_{kl}(n_1)  A_{ij}(m_2) B_{jl}(m_1) \cr
&& -{2 \over (m_1+m_2)^2}
A^\dag_{lj}(n_2)B^\dag_{ji}(n_1) B_{lk}(m_2) A_{ki}(m_1) \cr
&& \left({2 \over (n_1-m_1)^2}-{2 \over (m_1+m_2)^2}\right)
B^\dag_{kl}(n_2) A^\dag_{lj}(n_1)  B_{ki}(m_2)A_{ij} (m_1) \cr
&& \left. {2 \over (n_1-m_1)^2}
B^\dag_{ji}(n_2) B^\dag_{ik}(n_1) B_{jl}(m_2) B_{lk} (m_1) \right\} \label{ham2}\\
T_{3 \rightarrow 1}  & = &
\left({g_{YM}^2 N K \over{2}  \pi}\right) \sum \delta(n_1 - m_1 - m_2 - m_3) \times \cr
&&
\left\{-{2 \over (n_1-m_1)^2} A^\dag_{lj}(n_1) B_{lk}(m_3) A_{ki}(m_2)A_{ij} (m_1) \right.\cr
 &  &+ \left({2 \over (m_2 + m_1)^2} -{2 \over (n_1-m_1)^2}\right)
A^\dag_{lj}(n_1) A_{lk}(m_3) B_{ki}(m_2) A_{ij} (m_1)  \cr
&& + {2 \over (m_1+m_2)^2}
 A^\dag_{kl}(n_1)  A_{ki}(m_3)  A_{ij}(m_2) B_{jl}(m_1) \cr
& & -{2 \over (n_1-m_1)^2}
B^\dag_{ik}(n_1) A_{ij}(m_3) B_{jl}(m_2) B_{lk} (m_1)\cr
& & + \left({2 \over (m_1+m_2)^2}-{2 \over (n_1-m_1)^2}\right)
B^\dag_{kl}(n_1)  B_{ki}(m_3)  A_{ij}(m_2) B_{jl}(m_1) \cr
&& \left.+{2 \over (m_1+m_2)^2}
B^\dag_{ji}(n_1) B_{jl}(m_3) B_{lk}(m_2) A_{ki}(m_1) \right\} \label{ham3} \ .
\end{eqnarray}

At this point, a computer program must be written to generate the set
of states and the elements of the mass operator $M^2 = 2 P^+ P^-$.

As an example, for $K=5$ and $Q=1$, we find
\begin{equation}
T =
{g_{YM}^2 N K \over {2}\pi}
\left(
\begin{array}{ccccc}
 \frac{13}{8} & \frac{1}{8} & -\frac{1}{2} & \frac{1}{2} &
   -\frac{1}{4} \\
 \frac{1}{8} & \frac{13}{8} & \frac{1}{2} & -\frac{1}{2} &
   -\frac{1}{4} \\
 -\frac{1}{2} & \frac{1}{2} & 1 & 1 & 0 \\
 \frac{1}{2} & -\frac{1}{2} & 1 & 3 & 0 \\
 -\frac{1}{4} & -\frac{1}{4} & 0 & 0 & \frac{3}{2}
\end{array}
\right)
\end{equation}
whose eigenvalues in units of $g_{YM}^2 N /{2} \pi$ are
\begin{equation}
\{0,6.25,10,10,17.5\} \ .
\end{equation}

Strictly speaking, eigenvalues of just the $T$ without the
contribution from $V$ are unreliable since they correspond to taking
massless fermions as the matter fields. Nonetheless, it is encouraging
to find an exactly massless state in the spectrum which would survive
the limit of strong gauge coupling. The massless
states for these $m^2=0$ cases continue to be present as the values of $K$
are increased.

The actual computation reported in section \ref{sec:low} is the
computation of the spectrum of
\begin{equation} {M^2 \over m^2} ={1 \over m^2} (V + T) \end{equation}
where, for definitiveness, we set the dimensionless parameter
\begin{equation} x = {{ 2} \pi m^2  \over g_{YM}^2 N} = 10^{-3} \ . \end{equation}

The result of carrying out the calculation for
$K=5,7,9,11,13,15,17,19$, is summarized in table \ref{tablec}.  There,
we tabulate the calculated value of the hadronic bound state mass $M$ for
fixed $Q$ as $K$ is increased. The $K=\infty$ is a result of linear
extrapolation illustrated in figure \ref{extrapolate}. We observe
that the lightest bound state appears to be in the $Q=3$ sector, at
least among the fermionic states.  However, the quantity $M/Q$ which
determines $\mu_{crit}$ appears to be slowly decreasing
as $Q$ is increased.

\begin{table}
\begin{tabular}{|c|c|c|c|c|c|c|c|c|c||c|} \hline
 & $K=5$ & $K=7$ & $K=9$ & $K=11$ & $K=13$ & $K=15$ & $K=17$ & $K=19$ & $K=\infty$ & $M/Q$ \\ \hline
$Q=1$ & 4.60 & 5.08 & 5.42 & $\cdots$ &$\cdots$ &$\cdots$&$\cdots$&$\cdots$&6.42 & 6.42 \\ \hline
$Q=3$ & 3.42 & 3.67 & 3.85 & 4.00 & 4.10 &$\cdots$&$\cdots$&$\cdots$& 4.49 &  1.50 \\ \hline
$Q=5$ &- & 5.51 & 5.87 & 6.14 & 6.36 & 6.54 &$\cdots$&$\cdots$& 7.40 & 1.47 \\ \hline
$Q=7$ &- &- & 7.55 & 7.97 & 8.30 & 8.58 & 8.82 &$\cdots$ & 10.19 & 1.46 \\ \hline
$Q=9$ &- &- &- & 9.57 & 10.03 & 10.041 & 10.73 &$\cdots$ & 12.83 & 1.43 \\ \hline
$Q=11$ &- &- &- &- & 11.59 & 12.08 & 12.49 & 12.85 & 15.54 &  1.41 \\ \hline
\end{tabular}
\caption{Mass $M$ of the lightest fermionic bound state in the fixed
$Q$ sector for various $K$. The ``$-$'' indicates entries which are
not defined. The ``$\cdots$'' indicate entries which are well defined
but have not been computed due to limits in computational
resources. These are the data presented in figure
\ref{extrapolate}. Note that $M$ as a function of $Q$ is minimized at
$Q=3$. However, $M/Q$ as a function of $Q$ appears to slowly be
decreasing monotonically.\label{tablec}}
\end{table}

Similar computations can be carried out for the bosonic bound states
when the values of $K$ are taken to be even. We did not perform the
computation at the same level of precision for the bosonic bound
states as we did for the fermions. The plot analogous to figure
\ref{fullspec} is in figure \ref{fullspec_b} below. From figure
\ref{fullspec_b}, it is quite apparent that the pattern of states with
smallest $M/Q$ is becoming degenerate at around $\mu_{crit} = M/Q
\approx 1$ as $Q$ is increased.

\begin{figure}
\centerline{\includegraphics{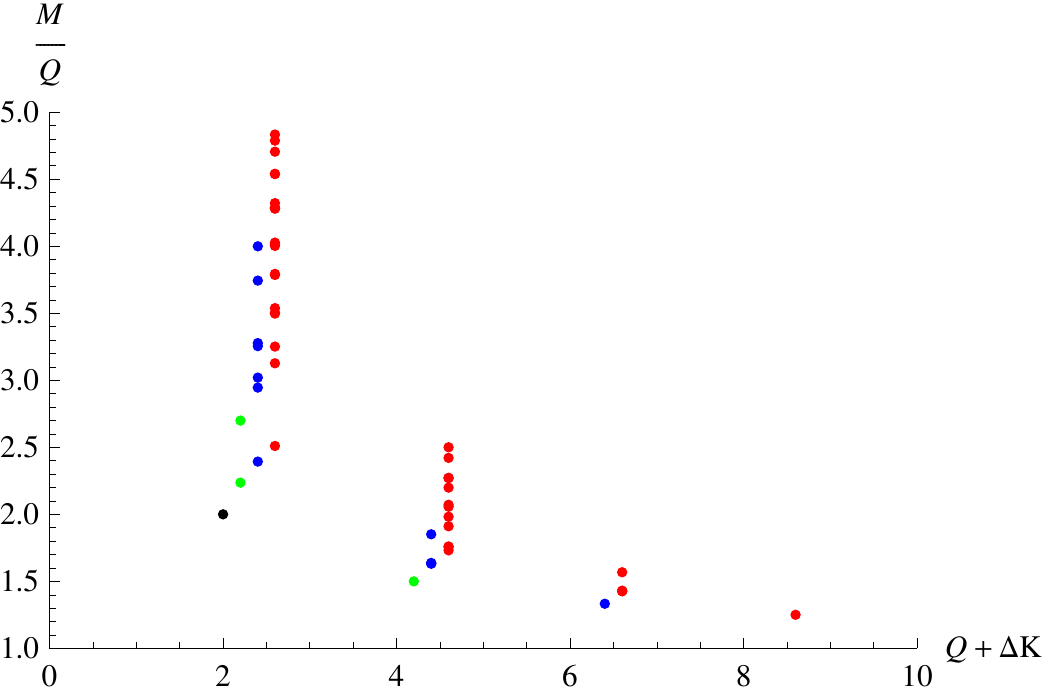}}
\caption{Spectrum of bosonic bound states for $K=4,6,8,10$.
This figure is the analogue of figure \ref{fullspec} for the fermionic bound
states.  \label{fullspec_b}}
\end{figure}

There is one additional feature which is notable regarding the bosonic
spectrum: the state with smallest $M$ is in the $Q=0$ sector. For
$K=4,6,8,10$, we find the masses presented in table \ref{Q0table}.

\begin{table}
\begin{tabular}{|c|c|c|c|c|c||c|} \hline
 & $K=4$  & $K=6$ & $K=8$ & $K=10$  & $K=\infty$ & $M/Q$ \\ \hline
$Q=0$ & 2.31 & 2.47 & 2.58 & 2.67 & 2.88 & $\infty$ \\ \hline
\end{tabular}
\caption{Mass $M$ of the bosonic bound state in the $Q=0$ sector for $K=4,6,8,10$. \label{Q0table}}
\end{table}

A closer examination of the wavefunction indicates that the lightest
state is mostly a mixture of the ``two bit'' states of the form
\begin{equation}
\sum_p c_p \mbox{Tr} A^\dag(p) B^\dag(K-p) |0 \rangle \ .
\end{equation}
This is analogous to what was found for the adjoint Majorana model
\cite{igor2}.  These are interesting features of our model from the
point of view of its dynamics at vanishing chemical potential.  As
should be clear from the right most column in table \ref{Q0table},
however, the $Q=0$ states do not have any impact on the physics at
finite chemical potential $\mu$.

\end{document}